\documentclass[fleqn,10pt]{wlscirep}
\usepackage[utf8]{inputenc}
\usepackage[T1]{fontenc}
\usepackage{mathtools}
\usepackage{times}
\usepackage{lineno}
\usepackage{xcolor}
\usepackage{xspace}
\usepackage{hyperref}
\usepackage{comment}
\usepackage{adjustbox}
\usepackage{graphicx}%
\usepackage{multirow}%
\usepackage{amsmath,amssymb,amsfonts}%
\usepackage{amsthm}%
\usepackage{mathrsfs}%
\usepackage[title]{appendix}%
\usepackage{xcolor}%
\usepackage{textcomp}%
\usepackage{manyfoot}%
\usepackage{booktabs}%
\usepackage{algorithm}%
\usepackage{algorithmicx}%
\usepackage{algpseudocode}%
\usepackage{listings}%
\usepackage{adjustbox}
\usepackage{ulem}

\newcommand{\new}[1]{\textcolor{black}{#1}}
\usepackage{ulem}
\usepackage{verbatim}
\title{Highly engaging events reveal semantic and temporal compression in online community discourse}
\author[1,2,+]{Antonio Desiderio}
\author[1,2]{Anna Mancini}
\author[1,2,*]{Giulio Cimini}
\author[3,4$\ddag$]{Riccardo {Di Clemente}}
\affil[1]{Physics Department and INFN, University of Rome Tor Vergata, Via della Ricerca Scientifica 1, Rome, 00133, Italy}
\affil[2]{Centro Ricerche Enrico Fermi, Via Panisperna 89a, Rome, 00184, Italy}
\affil[3]{Complex Connections Lab, Network Science Institute, Northeastern University London, 58 St Katharine's Way, London, E1W 1LP, United Kingdom.}
\affil[4]{ISI Foundation, Via Chisola 5, Turin, 10126, Italy.}
\affil[$\ddag$]{riccardo.diclemente@nulondon.ac.uk}
\affil[*]{giulio.cimini@roma2.infn.it}
\begin{abstract}
People nowadays express their opinions in online spaces, using different forms of interactions such as posting, sharing and discussing with one another. 
How do these digital traces change in response to events happening in the real world?
We leverage Reddit conversation data, exploiting its community-based structure, to elucidate how offline events influence online user interactions and behavior.
Online conversations, as posts and comments, are analysed along their temporal and semantic dimensions. 
Conversations tend to become repetitive with a more limited vocabulary, develop at a faster pace, and feature heightened emotions.
As the event approaches, the shifts occurring in conversations are reflected in the users' dynamics.
Users become more active and they exchange information with a growing audience, despite using a less rich vocabulary and repetitive messages.
The recurring patterns we discovered are persistent across a wide range of events and several contexts, representing a fingerprint of how online dynamics change in response to real-world occurrences.
\end{abstract}
\begin{document}
\flushbottom
\maketitle
\thispagestyle{empty}
\section*{Introduction}
In today's world of data, the detection of human interactions is increasingly being realized through the continuous stream of signals generated \cite{sapiezynski2019interaction,moro2022identify} and the knowledge extracted from them can be fed into reliable predictive models \cite{eagle2009inferring,lu2012predictability}, continuously refining our portrait of human behaviour \cite{lazer2009life,wagner2020science}. 
As human beings, we are social animals living in a community \cite{cherry1966human,emery2007introduction,littlejohn2010theories}
and we communicate social issues with others to share our ideas and views \cite{lee2013processes,kahne2018political,jin2020ll}.
Communication is a complex phenomenon, shaped by individuals' responses to external stimuli through various channels of communication \cite{stevens1950introduction,anguera10}.
Nowadays, online social networks represent the most popular means through which humans communicate
\cite{heidemann2012online,bennett2011collective}, digest 
information \cite{cinelli2020covid,menczer2018fake} and engage in discussions about offline events \cite{halu2013connect}.
These digital discussions provide an unprecedented amount of data that can lead to a quantitative understanding of how we interact with each other \cite{omodei_characterizing_2015} and, in turn help us address major socio-political challenges of our times \cite{lorenz2022systematic}.
For instance, by collecting tweets related to climate change conferences \cite{falkenberg2022growing} we can analyze the discussions and reveal a significant rise in ideological polarization due to the growing presence of right-wing activity.

Offline events such as political elections, championship sport matches, or large-scale epidemic outbreaks are characterised by a mass convergence of online attention and in turn these events can be used to \new{precisely} quantify  collective behavioral dynamics \cite{szell_contraction_2014,lorenz2022systematic}.
The research literature typically characterizes users' attention as the amount of engagement with news.
We understood that news propagates and fades away with a stretched-exponential law \cite{wu2007novelty}, using the news' popularity index submitted by users on Digg.com, and \new{characterized the} burst of activity followed by power-law relaxation using views of new Youtube videos \cite{crane2008robust,yang2011patterns}. 
Analyzing user-generated content and real-time communication platforms such as Twitter and Yahoo Research,
\new{we have measured} that temporal patterns in users' tweet streams changes from the baselines during shocking events \cite{sasahara2013quantifying,he2017measuring},
such as terrorist attacks, natural disasters or elections.
However, human interactions around social issues consist of a continuous dynamical exchange of ideas typical of communities \cite{candia2019universal}, defined as group of individuals who share common interests, characteristics, and interact with one another on a regular basis.
How do events affect community discourse, interaction frequency, and writing style?
Are these changes, along the temporal and semantic dimensions, specific to the type of event, or are they recurrent across events?
Offline events shape the environment in which online conversations take place, thereby changing the direction and altering the dynamics of online discussions, influencing how individuals digest online information \cite{lorenz2022systematic}.

We address these questions using Reddit conversation data. 
Reddit is a public online forum whose users interact with each other by submitting new posts and adding comments to existing posts or comments, thus creating conversation threads \cite{choi_characterizing_2015}. 
The wide thematic spectrum of Reddit conversations enables us to deepen our comprehension of human communication within communities \cite{proferes2021studying}: for instance it was shown how users become more intuitive and express their sadness during COVID-19 warning and lockdown phase \cite{ashokkumar2021social} or how users try to shift the point of view of their interlocutors according to their preconceptions \cite{monti2022language}.
We investigate the temporal and semantic dimensions of online community discourse during highly engaging events \cite{sasahara2013quantifying,he2017measuring,szell_contraction_2014}.
\new{
By examining changes in the dynamic structure and content of these discussions, we gain insights into community responses to key events, with implications for how information spreads and how public discourse is shaped in response to such events.
}
By analyzing the time sequence of comments, we can identify variations in discussion activity speed, while the semantic dimension uncovers unique patterns of words and statistically significant expressions within the conversation.
\new{
Shifts in community engagement marked by semantic redundancy and increased activity frequency, reflect the intensity and dynamics of collective responses to significant real-world events.
}
Reddit conversations during these events indeed display extreme variations: the frequency of replies increases, conversations develop at a faster pace and are repetitive, the use of word combinations changes, and there is an increase of total emotions shared.
\new{
These conversations evolve as users exchange messages around the event, reflecting heightened engagement.
}
Users express their opinions and thoughts on a given event through a comment, that is shared with the community at a specific \emph{time} and with a \emph{semantic} fingerprint. 
\new{
During events, users begin to increase their activity frequency, while also interacting with more users. 
High frequency of activity involves a lack of diversified language}, and extremely repetitive messages.
As users interact with a growing audience, joining the debate and \emph{de facto} broadening the exchange of information.
The semantic diversity of a user's conversation peers increases as they occupy a larger semantic space \cite{gonzalez2008understanding,lombardo2022mobility}, shifting the dialogue in practice.
\new{
The resulting picture reveals that the increased production of community content around offline events is accompanied by  semantic redundancy among users, which emerges alongside high activity frequency.
By dissecting communication dynamics in online communities, our approach enhances content moderation efforts to track the evolution of discourse over time, shedding light on how digital platforms function as spaces for collective knowledge-sharing and social cohesion in an increasingly online world.
}

\section*{Results and Discussion}

The Reddit platform consists of a vast collection of communities, each dedicated to a specific topic \cite{olson2015navigating}.
Here, we focus on communities with a large user base that discuss U.S. politics (r/politics) and European politics (r/europe), as well as U.S. basketball (r/NBA) and football (r/NFL). 
Our Reddit dataset comprises over 60 million comments, with a time range spanning from January 01, 2020 to January 31, 2021. 
This period includes a broad range of events such as the COVID-19 pandemic, the U.S. 2020 elections, NBA interruption, Kobe Bryant death, several NFL matches, etc...(see Supplementary Information, Section 1 for the full list of the events considered).
\new{
We identified notable events on a weekly basis using a fixed time window, which allowed us to include events spanning multiple days while ensuring reliable statistical analysis of highly engaging events through aggregated signals.
Details on the criteria and definitions can be found in the following section and Supplementary Information, Section 2.
}
\subsection*{Burst of Activity and Conversation Characterization.}

\begin{figure*}[!ht]
	\centering
	\includegraphics[width=1\textwidth]{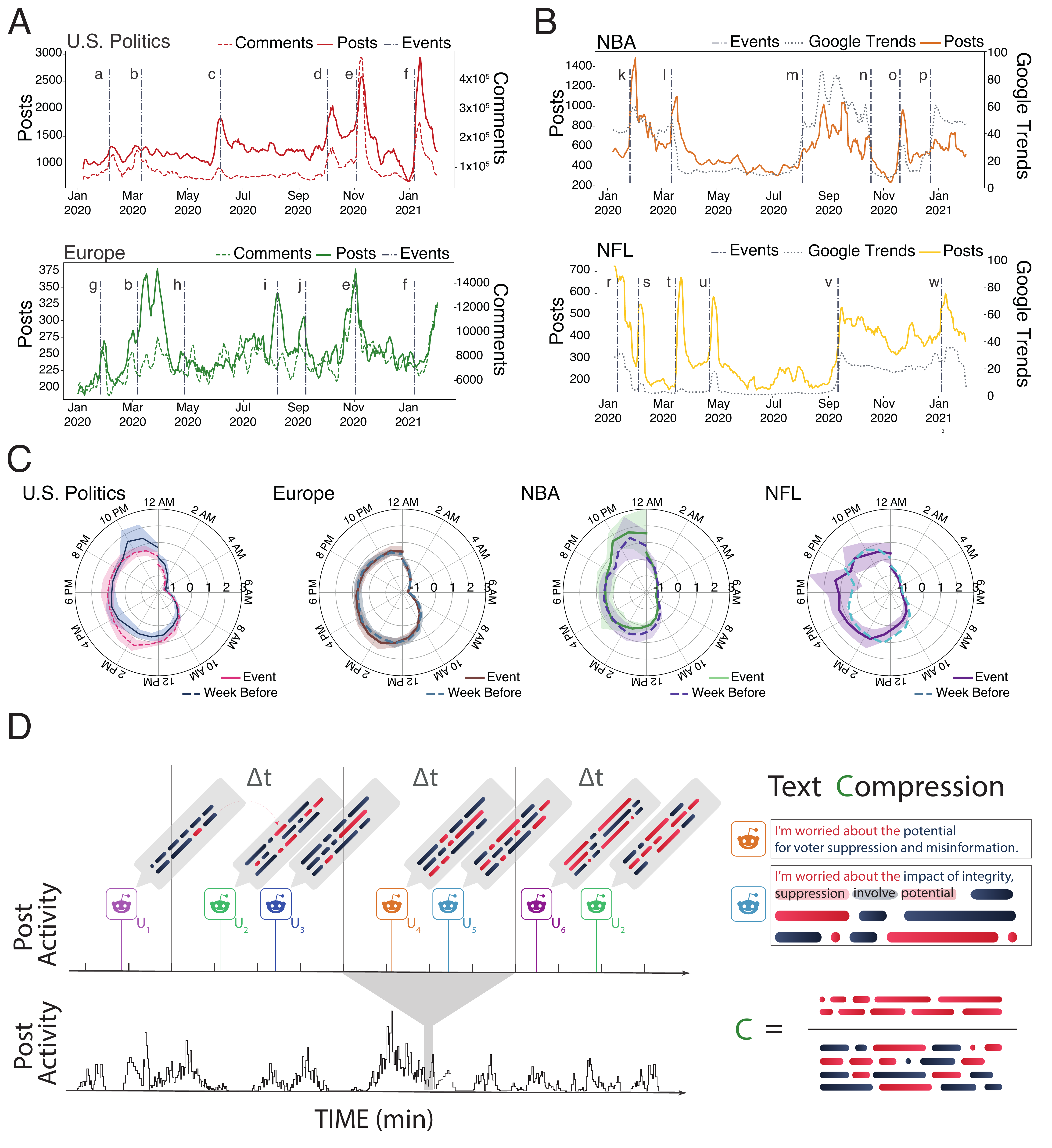}
	\caption{\textbf{Burst of Activity and Conversation Characterization}
		In subplots A-B we apply a 7-day moving average to the time series.
		A) Number of posts (solid line) and comments (dashed lines) for the U.S. politics community (upper panel) and European community (lower panel). The grey vertical dashed-dotted lines mark the highly engaging events and correspond to the peaks of the signals.
		B) Number of posts compared to the Google Trends for the NBA community (upper panel) and NFL community (lower panel).
		C) Radar plots showing, for each subreddit, the average Z-scored hourly activity in the week before (dashed line) and that of the event (solid line), with the shaded area representing the standard deviation. 
        For the European community, the Amsterdam timezone is selected, while the US/Eastern timezone is employed for all other communities.
        D) Schematic representation of how we characterise a conversation. For each post we capture the temporal dimension as the time series extracted by counting the comments underneath within a $\Delta t=5$ minutes time interval, and the semantic dimension by merging all the comments into a single text, whose compression is obtained as the ratio between the number of unique patterns of words (in red) and of all words (unique and repeated).}
    \label{fig1}
\end{figure*}

A burst in the overall conversations' volume around an event is the hallmark of its attractiveness \cite{wu2007novelty,sasahara2013quantifying}.
\new{
We identify peak weeks of heightened engagement by ranking weekly bursts based on daily z-score variations in posts (see Supplementary Information, Section 2 and Methods).
To contextualize these peaks, we use nearby events taken from Wikipedia pages (see Supplementary Information, Section 1, Table S2 for the pages retrieved). We note  that certain events -- such as COVID-19 in the U.S. and the Capitol Hill incident in Europe -- appear with shifts in time across geographic areas, likely due to delays in the public's response timing.
}
We trace more than 20 highly engaging weeks in the chosen Reddit communities, as documented on the Wikipedia pages.
On the contrary, we did not considered in our analysis events such as the first impeachment trial of U.S. president Donald Trump (Jan 16, 2020), and Bulgarian protests in July 2020 -- listed on the Wikipedia pages -- since they received limited engagement from the respective communities on Reddit.
Figure \ref{fig1}A shows the burst of activity within Reddit political communities, in terms of overall number of daily posts and comments generated around highly engaging events.
In general, volumes of both posts and comments increase during the event, with some noteworthy exceptions (e.g. COVID-19 for the U.S. politics community where comments grow much more than posts). 
To cross-check the events selected we have integrated into our analysis Google Trends data (using \emph{nba} and \emph{nfl} as query terms for r/NBA and r/NFL, respectively).
Figure \ref{fig1}B shows how the time series for the number of posts of the sport communities and the Google Trends are strongly correlated (NBA Correlation 0.7, NFL Correlation 0.8), and the peaks mostly coincide, meaning that people search for events (Google Trends) as they talk more about them (Reddit).
For these communities, we observe events that span several weeks, coinciding with weekly matches, and in this case, we consider the start and end of this period as an event.

To gain a more comprehensive understanding of the interplay between external events and the communication patterns within the Reddit communities, a common approach is to explore the users' behavior around the observed peaks \cite{he2017measuring,sasahara2013quantifying}.
Figure \ref{fig1}C displays the Z-scored hourly activity of the week before and that of the event. 
We observe that the digital circadian rhythm of content production is different around specific hours, most likely modified by the events \cite{sasahara2013quantifying}. 
For instance, in sports-related events, the difference is most noticeable around the match kickoff and the end of the game.
In the EU politics case we do not observe a relevant gap, while in the U.S. politics case there is a marked shift in the evening. 

\new{Given that offline} events influence users' online activity \cite{sasahara2013quantifying,lorenz2019accelerating,fortier2017hypermodern,malley2008some}, how do these events change the way people communicate with each other?
On Reddit, users discuss about specific topics or events by writing posts and commenting to other posts or comments. Hence, we can consider a post and all its underneath comments  as a single conversation.
Figure \ref{fig1}D illustrates how Reddit posts are depicted and compared along the temporal and semantic dimensions.
From each post we extract a time series by considering time intervals of length $\Delta t$, starting from the creation of the post, and counting how many comments are written within each of these intervals.
We also extract a text, or document, for each post by joining all the comments underneath, and compute its compression as the fraction of unique words to the total number of words in the document.

\subsection*{Accelerating online conversational pace during offline events.}

We measure Dynamic Time Warping (DTW) and Coherence distance between conversations of one week and the week before to capture the temporal shift of conversation dynamics (see Methods). 
The aim of DTW \cite{berndt1994using} is to find the optimal alignment between two time series by warping one of them in a nonlinear way. 
This alignment process captures stretching and compression between the two series. 
When the DTW distance increases, the two series become less similar (temporal mismatch/distortions), as the alignment process requires more warping or compressing of the sequences to work properly.
Coherence, instead, allows to measure possible enhancing time series' relationships between weeks by computing the frequency spectra, and detecting common frequency patterns \cite{grinsted2004application,maharaj2010coherence}. 
The purpose of coherence is to measure the degree of linear synchronization between time series, providing insights into how much two time series are correlated at different frequencies and it indicates how well the phases align at different frequencies (consistency of their temporal shifts).
Conversely, low coherence values point to inconsistent or random temporal shifts (see Methods).
\new{
Coherence increases significantly during event weeks compared to baseline periods (weeks without events). This increase reflects a structural shift in conversations, which can either enhance (constructive) or disrupt (destructive) their dynamics.
}

\begin{figure*}[ht!]
	\centering
	\includegraphics[width=\textwidth]{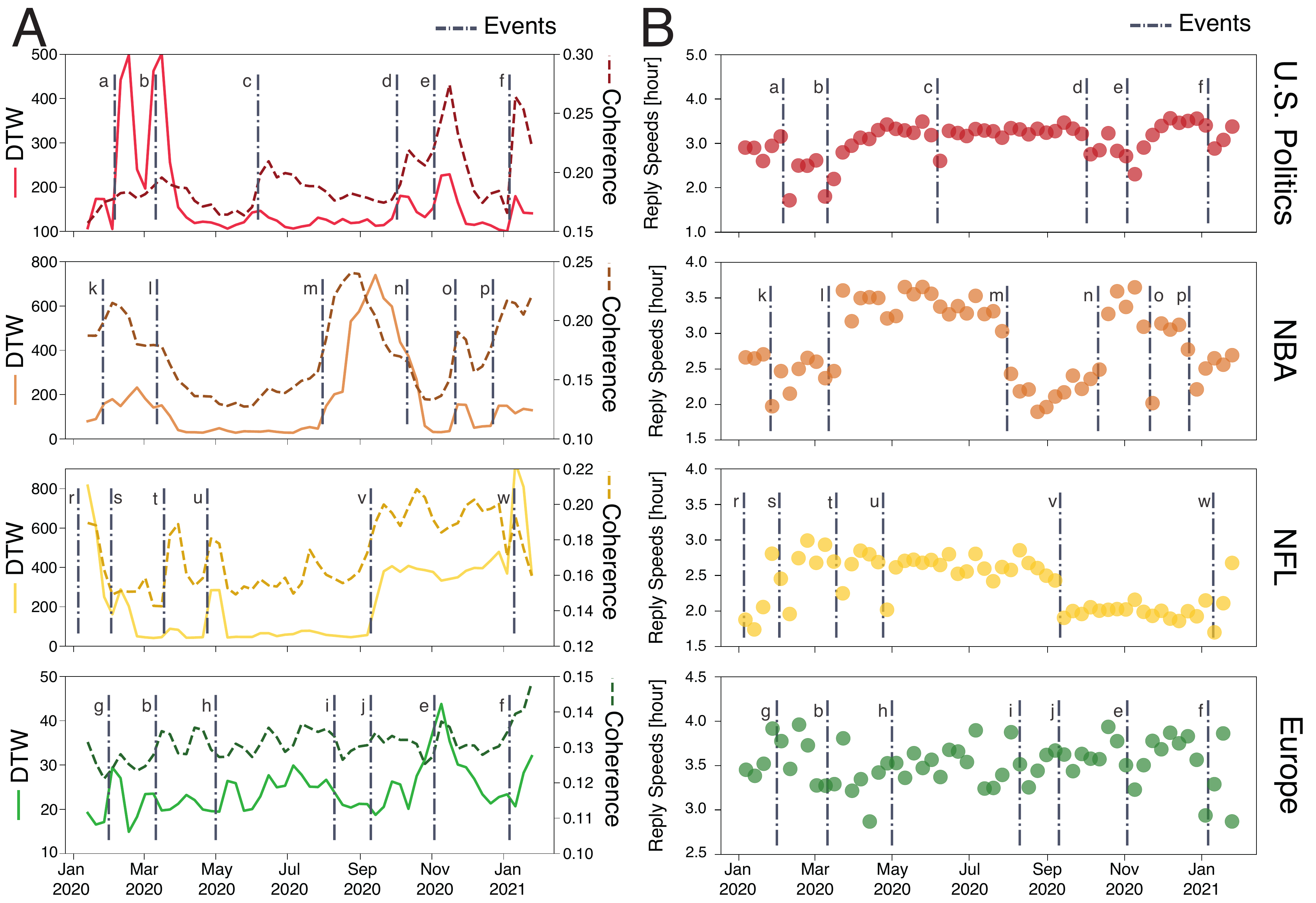}
	\caption{\textbf{Temporal Dimension.}
		A) Average Dynamic Time Warping (solid) and coherence (dashed) distances between the conversations of a week and of the previous one, for each subreddit. 
		B) Average reply speeds, for each subreddit. In all panels, the grey vertical dashed-dotted lines mark the events.
	}
	\label{fig2}
\end{figure*}

\new{
Figure \ref{fig2}A illustrates the average distances of DTW and coherence across events. For most of the analyzed events, significant changes are evident, with an average variation exceeding $39\%$. However, variations differ across events and communities. 
For instance, sharp spikes are observed during COVID-19 discussions in U.S. politics and the NFL Playoffs in January 2021.
In sports-related cases, large variations occur at the start of tournaments, but the average distances stabilize once the tournament progresses. 
Notably, events like the NFL and NBA show identical variations, both exceeding $80\%$. 
}
Meanwhile, the European community displays a marked variation \new{of 35 \%} only for the U.S. 2020 election.
\new{
Overall, coherence and DTW variations display consistent patterns, indicating shifts in conversation dynamics. 
However, coherence values remains consistently lower (below 0.35) than DTW, suggesting that while events affect conversation flow, shifts are not entirely synchronized in frequency and phase.
This implies that DTW detects more immediate and intense shifts, while coherence highlights more gradual, frequency-based alignment in conversational patterns.
}
Since DTW is sensitive to time distortion and different speeds, we validate the results by testing against null models obtained through randomization of the timestamps of the comments to statistically validate the changes in the way conversations are structured along the temporal dimension (Supplementary Information, Section 3).

Another interesting quantity to look at is the reply speed, defined as the temporal distance between a comment and its response. We find that the weekly distributions of reply speeds are well approximated by Log-Normal distributions, in agreement with other analyses of human temporal patterns \cite{kaltenbrunner2007homogeneous}.
We observe that there is a decrease of the reply speed during the events (see Figure \ref{fig2}B), exceeding 30\% in the majority of cases: the peaks of the distributions during the events are getting sharper and shifting to lower values (see Supplementary Information, Section 4 for the standard deviation of the reply speeds).
These variations are not the same for all the events, due to their heterogeneous attractiveness.
Furthermore, we observe that the reply speed remains consistently low -- approximately at the same level -- for events that span several weeks, such as NFL matches or the initial lockdown period in Europe.
We can conclude that, during highly engaging events, conversations along the temporal dimension are structured differently and take place with an overall faster pace.

\subsection*{Reduced focus yet increased diversity in conversation.}

\begin{figure*}[ht!]
	\centering
	\includegraphics[width=1\textwidth]{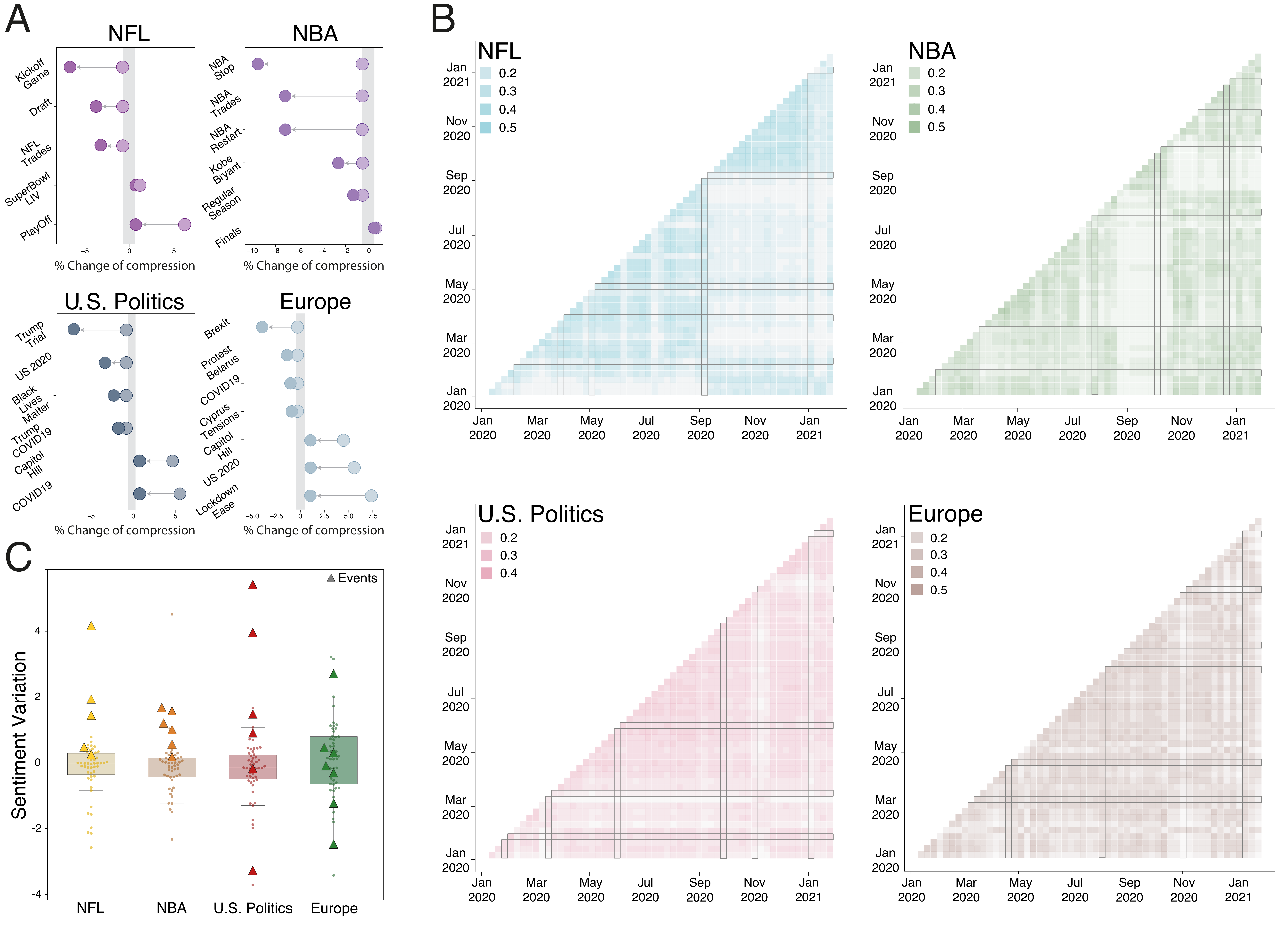}
	\caption{\textbf{Semantic Dimension.}
		A) Percentage change of compression between the week associated with the event (darker) and the week before (lighter) for each subreddit. The grey shaded vertical area is the standard deviation of the mean change between one week and the preceding week.
		B) Jaccard index among statistically relevant bi-grams between all weeks, the lighter the color the more the weeks are dissimilar. Events are marked with grey lines.
		C) Emotion variation for each subreddit between consecutive weeks. The triangles mark the variation associated to the events.
	}
	\label{fig3}
\end{figure*}

To measure the focus of conversations on a specific topic, we explore the information content of the text associated to each conversation thread.
We measure the compression of the conversation using the Lempel-Ziv complexity index, which measures the repetitiveness of the content (see Methods for further details). 
The idea behind Lempel-Ziv complexity is to measure the amount of information in a sequence or a string of symbols by identifying and encoding repeated patterns. 
Figure \ref{fig3}A shows the compression's variation between the week of the event and the week before: a negative variation outlines that conversations have become more repetitive. 
\new{
The grey shaded area represents the standard deviation of the variation in compression, which is generally close to zero, indicating stability over time, with spikes corresponding to highly engaging events.
}
Most of the events are characterised by a large variation in terms of compression level with respect to the preceding period; however, while the discussions about sports become in general more repetitive, the political discussions tend in the opposite direction.
\new{
For instance, events like the NBA Finals and NFL Kickoff show a variation in compression lower than $-6\%$, indicating an increase in repetitive content as users converge on specific recurring topics within the week.
Conversely, political events such as Capitol Hill and Lockdown Ease demonstrate positive variations, with changes exceeding +3\%, suggesting an evolving discourse during these periods.
}

Compression, however, focuses only on words, while people tend to repeat certain structures such as word sequences or phrases, which can be important for conveying meaning or establishing a sense of belonging of a user to the community, especially during a particular event.
To capture changes in language before and after events, we detect the statistically significant structures within conversations. 
We generate an ensemble of document realizations for each week by randomizing the order of words, and compute the relevant bi-grams against the ground truth to assess their statistical significance.
We limit our analysis to the top bi-grams and we exploit them to compare the weeks using Jaccard similarity index among bi-grams (see Figure \ref{fig3}B and Methods). 
\new{
For most events, regardless of the topic, Jaccard similarity index values remain below 0.3 when comparing event-related weeks with other weeks. 
This indicates the presence of distinct statistically relevant bi-grams during event weeks.
In sports events, match weeks consistently exhibit a Jaccard similarity index above 0.4 when compared to each other, but below 0.25 when compared to the other weeks.
This generates distinct clusters of linguistically similar weeks, visible in Figure \ref{fig3}B as areas with similar Jaccard index values. 
A comparable pattern emerges during the U.S. 2020 election weeks (October 2020), where Jaccard indices remain above 0.36, reflecting a clustering of linguistically similar weeks. This linguistic consistency aligns with observations from the temporal analysis.
}
We derive the dissimilarity index from the Jaccard indices, which measures the number of weeks where the Jaccard index falls below the median value for a given week (see Supplementary Information, Section 5).

We further perform sentiment analysis to provide a more complete understanding of conversation content and of people's perceptions and attitudes towards an event. 
Sentiment analysis is a standard technique in online social network analysis to capture the polarity of a text \cite{box2021meaningful,bovet2018validation,matalon2021using}.
First, we compute the sentiment of each post and comment using VADER. Sentiment varies between -1 (negative) and +1 (positive). 
We binned this interval and compute, for each week, the histogram of post/comment sentiment values. 
Then, we compute the Z-score of each bin by using the average value and standard deviation of all weeks. 
Finally, we compute the variation of the emotion between a week and the week before.

Generally, there is a consistent positive emotion shift \new{of more than 1.2 standard deviation} between the week of the event and the preceding week, compared to the variation observed between two consecutive weeks prior to the event (see Figure \ref{fig3}C).
As in the previous results, all emotion changes for the weeks of NBA and NFL matches lie on the upper tail of the distribution. Meanwhile, in the U.S. politics community we find significant variations for the election weeks and the entire Black Lives Matter protest period, while in the EU case during the first COVID-19 lockdown. 
\new{
Overall, we observe consistent sentiment variations within topics, such as between NFL and NBA, while noting dissimilarity across different categories, particularly between U.S. politics and the NBA.
Notably, this result aligns with the compression analysis, suggesting a consistent pattern within topics, where highly engaging events lead to increased linguistic predictability and shared emotional structures in communities.
}
We can conclude that communities express their views and feelings towards the event in a multifaceted manner, with large variations in sentiment and expressions defined by different combinations of words.

\subsection*{User dynamics reveal amplified repetition along with heightened speed.}

When people engage in a conversation, they exchange comments with one another and within the community, giving rise to a dynamic process of communication.
The dynamical changes of conversations as a whole due to the occurrence of a particular event, that we observed in the previous results, also imply the existence of shifts in temporal activity and semantic structure at the level of individual users. 
Hence, in this section, we focus our analysis on individual behaviors. 
Events typically involve an higher number of users who are active solely during the event. 
As a result, the observed conversational shifts in the previous sections -- even if validated with null models -- may be attributed to these random users \cite{szell_contraction_2014}.
Consequently, in this analysis, we consider only recurrent or dutiful users -- those who actively and consistently engage in the community over several weeks (see Methods for details).
Furthermore, we consider an additional dimension given by the number of conversation peers of each user, that is, how many neighbors she has in the network of social interactions.

\begin{figure*}[ht!]
	\centering
	\includegraphics[width=1\textwidth]{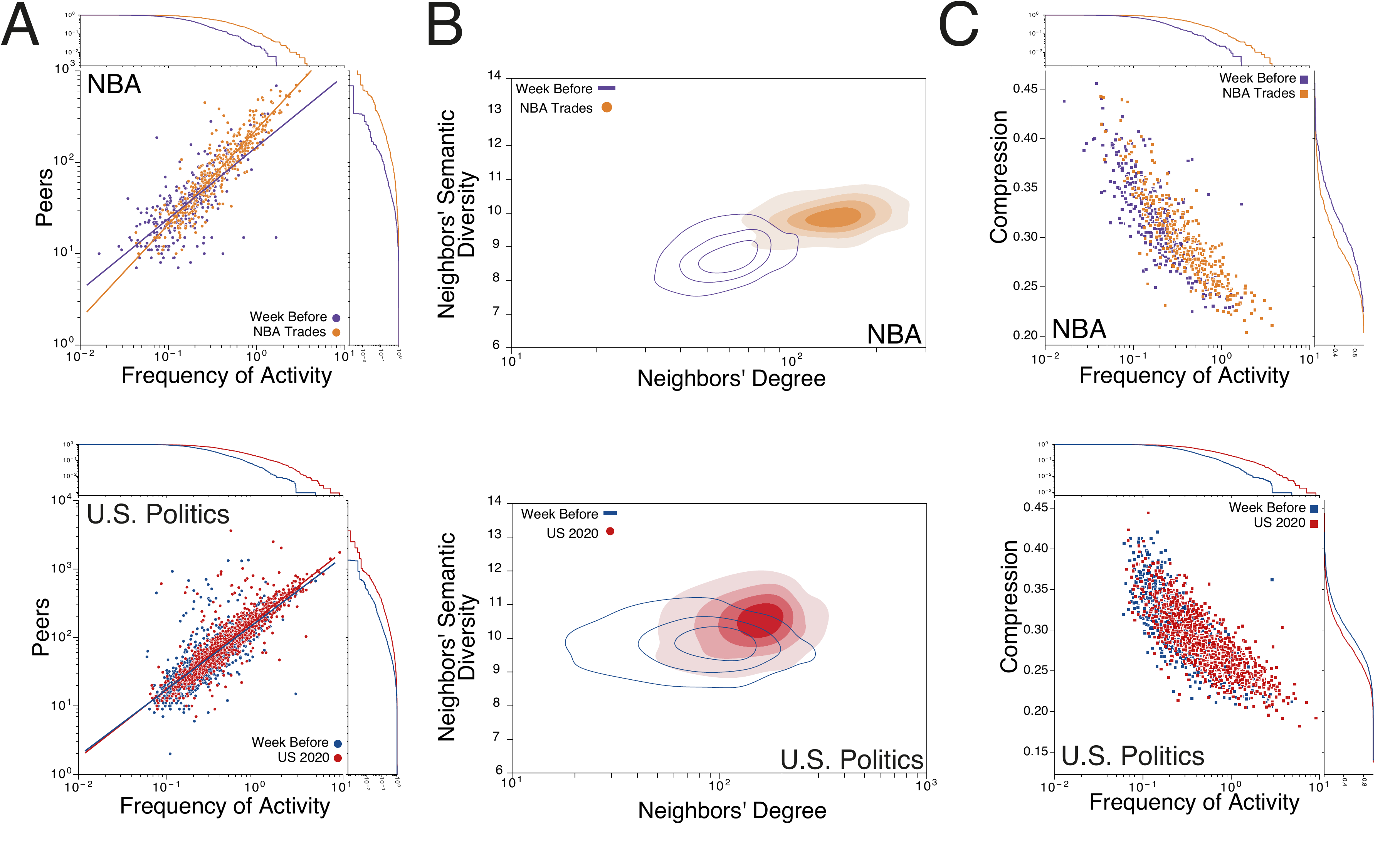}
	\caption{\textbf{Users' dynamics.}
		In the following subplots the data used on the left panels are of the users active on the subreddit r/NBA during the NBA Trades, while on the right of the users on r/politics during the U.S. 2020 election.
		A) The central panels show the relation between the frequency of activity of each user and the number of interacting peers (the degree). The marginal plots report the survival function of each variable for the two weeks.
		B) The density plots show the variations of the peers' degree and semantic diversity.
		C) The panels show the relation between user's compression and frequency of activity. Marginal plots report the survival function of each variable for the two weeks.
	}
	\label{fig4}
\end{figure*}

We characterise the individual temporal dimension using the frequency of activity, considering both comments and posts contributed by each user, as an indicator of her level of \new{time-based} engagement with the community.
We find that there is a \new{power-law} correlation between the users' activity frequency and the number of users with whom it interacts, the users' degree, regardless of the event ($R^2 > 0.7$, see Figure \ref{fig4}A).
\new{
The relationship is sublinear, with an average exponent of 0.8 across all weeks, indicating that as users increase their activity frequency, they tend to interact with a disproportionately smaller number of peers.
However, the exponent increases during events, reflecting broader engagement with more users while still maintaining sublinear growth (See Supplementary Information, Section 6, Table S6).
}
Hence, during an event, users tend to increase their activity frequencies and they engage in conversations with an expanding group of users (see also the distributions shift in Figure \ref{fig4}A and Supplementary Information, Section 6 for more examples). 
As more users join the discussion surrounding the event, they become more engaged and reach a larger audience.
\new{
We analyze user dynamics to track changes in activity frequency and degree, providing a complete picture of their engagement. 
We use the Wasserstein distance \cite{kantorovich1960mathematical} to compare distributions with varying supports and capture dynamic shifts.
}
We consider the two political communities during the shared events (U.S. 2020 election and the Capitol Hill riot) and we observe that there are no changes (Wasserstein distance near zero) across events in the European case, contrary to the U.S. case \new{(Wasserstein distance $>0.3$)}, showing that the users' engagement level is not simply related to the community volume production (see Supplementary Information, Section 6 and 7).

We then move to the analysis of the semantic dimension, considering all the posts and comments contributed by a user in a given week. 
We found that at the conversation level the combinations of words chosen by users to express their feelings changes during the events (Figure \ref{fig3}).
By mapping the text of comments into the Mikolov semantic space \cite{mikolov2013distributed}, where words that share similar contexts in the corpus are located in close proximity, we can capture users' movements in the conversation by measuring their semantic diversity (see Methods). 
Due to the shorter text data at the individual user level, statistically validated bi-grams can be noisy in capturing semantic diversity.
We find that during events, users' peers become more semantically dispersed, and connected at the same time, as shown by the shifts in Figure \ref{fig4}B. 
\new{
Notably, semantic diversity tends to increase during events, with high values exhibiting further growth while maintaining a stable spread (See Supplementary Information, Section 6, Table S6).
}
Other events can be found in Supplementary Information Section 6 and 7. 
Additionally, we find that the average semantic displacement of each post, defined as the average distance in the semantic space between a comment and the succeeding one, tends to increase as the semantic diversity of the user also increases (see Supplementary Information, Section 8).
\new{
We observe no difference in post displacements between the communities; conversely, NBA and NFL cluster at lower semantic diversity values, U.S. Politics exhibits the highest variability and range, and Europe overlaps with the other political community.
}
In other words, as the users' peers become more semantically dispersed and connected, they are introducing new and varied semantic structures into the conversations.
Finally, we notice that as the users' activity frequency increases, their semantic compression also grows (Figure \ref{fig4}C).
\new{
We confirmed that this trend is not solely due to text size growth, as predicted by Heaps' Law, which predicts that larger texts introduce fewer new words (see Supplementary Information, Section 9). 
The observed increase in semantic compression and activity frequency persists across text sizes during highly engaging events, indicating shifts in communication structure.
}
Overall, during events of highly engaging events, users tend to interact with a greater number of peers and the messages they exchange become even more repetitive. 

\section*{Conclusion}

The fingerprints extracted from the digital discussions on Reddit provide evidence of how offline events are perceived by online users.
The increased production of online content regarding in-person events is characterized by discussions marked by semantic redundancy, which develop over time at an accelerated pace, regardless of the event type.
The observed changes in online social media discussions are reflected in the dynamics of users, where their semantic spaces shrink as activity frequencies rise.
By examining the language used by each user's peers, we discover that individuals with broader vocabularies engage more frequently, hence influencing the direction of the conversation. 

Our framework for evaluating the impact of offline events on the digital discourse of a community is subject to certain constraints.
\new{
First, using Reddit -- an online social network where users communicate anonymously, predominantly in English, and which is not a mainstream platform -- limits the range of our findings. 
The development of conversations can be influenced by various factors, including the nature of the topic, the language employed, and the characteristics of the participants \cite{stans2017role}.
To overcome the issue of topic specificity, we focused on various US communities with large user bases, such as politics, NBA, and NFL.
Conversely, anonymity may encourage users to express more genuine opinions, as they are less constrained by social consequences \cite{bernstein_4chan_2021}.
Users who engage in specific discussions may not represent the broader population, as their participation is shaped by the unique context of each thread.
To overcome this issue when analyzing user dynamics, we filtered out random users who interacted solely because of the events.
}
Second, we have selected the timeframe of 2020, which was characterized by a massive increase in the usage of online platforms by individuals due to the COVID-19 containment measures \cite{ashokkumar2021social}.
Given this limitation, we have explored the timeframe of June 2016 specifically for NBA and U.S. politics, and the findings substantiate our main results (refer to Supplementary Information, Section 10).
Finally, our results are derived from a single social media platform, Reddit, due to its community structure. 
\new{Yet our framework}, which relies solely on semantics and temporal aspects of online interactions, can be readily applied to other platforms, thereby corroborating our findings.

Our analysis contributes to a deeper understanding of how offline events are discussed by online communities \cite{candia2019universal}.
The dissemination of knowledge and the consumption of news \cite{del2016spreading,gravino2022supply} are crucial aspects of modern societies \cite{watts2021measuring}, fostering social cohesion by providing a shared awareness of nowadays events and promoting exchange of perspectives.
With advancements in technology, news receive more collective attention but individual exposure is shortening \cite{lorenz2019accelerating} and individual daily activity is more fragmented \cite{sullivan2018speed}.
Here, we explore the semantic component of online debate, demonstrating that semantic redundancy is always coupled with higher activity frequency. 
This could result from users repeatedly expressing the same concept at a higher frequency, thus not fostering much deeper conversation.
\new{
A certain degree of variability is observed across individual communities, but this is resolved when communities are grouped by topic (e.g., sports and politics).
This pattern suggests that the topic under discussion may be useful in further characterizing user behaviors.
}
Studying semantic recurrences over longer time scales can reveal how language and culture change and adapt over time \cite{fortier2017hypermodern}, which can have valuable implications for fields such as linguistics \cite{graddol2004future} and anthropology \cite{malley2008some}. 
Furthermore, our framework holds the potential to identify events and remove biases within corpora employed for ML pipelines, specifically by identifying and excluding event-related data \cite{meade_empirical_2022}.

\section*{Methods}

\subsection*{Dataset}
We retrieved Reddit conversation data from Pushshift \cite{baumgartner2020pushshift}, an API that regularly copies activity data of Reddit and other social media. We queried the service to retrieve information about the chosen subreddits' posts and comments from January 01, 2020 to January 31, 2021. The datasets was cleaned by removing posts/comments made by users with username ending with \emph{bot} and \emph{AutoModerator} (see Supplementary Information, Section 1, Table S3).
Google Search engine data were generated by the Google Trends platform and were retrieved via the Python package \emph{pytrends} (see Supplementary Information, Section 1, Table S2 for the keywords used).
Highly engaging events are selected according to the daily Z-score \new{variations of the Reddit posting activity}, where the mean $\mu_s(t) = \frac{1}{t} \sum_{t' = 1}^{t} x_s(t')$ and variance $\sigma^2_s(t)=\frac{1}{t} \sum_{t' = 1}^{t} \left[ x_s(t') - \mu_s(t) \right]^2$ of the time series $x_s(t)$ are used to compute it \cite{mancini_self-induced_2022}.
This quantity captures the variation of engagement of the community in a given week, thereby allowing us to rank weeks according to it.
The daily Z-scores \new{variations} are reported in Supplementary Information, Section 2.
The events were contextualized with Wikipedia by manually inspecting the corresponding page of the subreddit \new{and matching with the bursts} (see Supplementary Information, Section 1, Table S2 for the pages). The events considered for each community are reported in Supplementary Information, Section 1, Table S1.
\subsection*{Temporal Analysis}
To compute the hourly activity, we have first counted the comments/posts for each hour within a week, then we have computed the hourly Z-score with respect to the average hourly activity of the overall period.
We have extracted a time series from each post by considering time intervals of length $\Delta t$, starting from the creation of the post, and counting how many comments are written within each of these intervals. We consider a post lifetime of 24 hours and discarded comments written afterwards (less than 5\% of the total, on average). 
For each week we have considered only the top 100 posts by number of comments (accounting for over 50\% of comments) and we measured \emph{Dynamic Time Warping} (DTW) distance \cite{bellman1959adaptive} and Coherence between all the possible combinations of conversations of one week and the week before. 
Coherence has been computed via Welch's method \cite{welch_coherence} using Hann window, with an overlap of 50\% between the two time series \cite{virtanen2020scipy}. If we have two time series, $y(t)$ and $x(t)$, that are linked by a convolution relation and additive white noise $w(t)$ such that $y(t) = {H\otimes x}(t) + w(t)$, we can compute coherence as follows
\begin{align}
	C_{xy}(\omega) &= \frac{|S_{xy}(\omega)|^2}{S_{xx}(\omega)S_{yy}(\omega)}=\nonumber\\ 
    &=\biggl(1+\frac{S_{ww}}{S_{xx}^2|H|^2}\biggr)^{-1}=\begin{cases}S_{ww} \gg S_{xx}^2|H|^2 \implies C_{xy} \sim 0 \\
		S_{xx}^2|H|^2 \gg S_{ww} \implies C_{xy} \sim 1
	\end{cases}\,.
\end{align}
where $S_{xy}$ is the cross-spectral density between x and y, and $S_{xx}$ the auto spectral density (same for $y$).
If coherence increases, then the impulse response function $H$ is greater than white noise $w$; this means that the variability of $y$ can be well explained by the variability of $x$.
DTW is a technique mainly used to find the optimal match between two time series with different lengths by non-linearly mapping one signal to the other \cite{berndt1994using}.
The key idea is to create a matrix $M_{ij}$, where the entries are the distances between each point $i$ in the signal $x(t)$ and each point $j$ in the other signal $y(t)$. The matrix $M_{ij}$ can be interpreted as the weighted adjacency matrix of a graph, where the point $i$ is connected to the point $j$ with a weight $M_{ij}$. 
We can use the Dijkstra's algorithm to find the weighted shortest path through the graph (cumulative distances between each point), which corresponds to the optimal DTW path between the two time series \cite{berndt1994using}.
The reply speeds have been computed as the elapsed time between a comment and its response, in this case all the comments have been considered within a week.
\subsection*{Semantic Analysis}
For each post, we joined all the (lower-cased text of) comments underneath, respecting their temporal order, to obtain a document.
For each week we have considered only the top 100 posts by number of comments and we have computed the Lempel-Ziv complexity index \cite{lempel1976complexity}. 
The algorithm works by scanning a string sequence and identifying repeated patterns or substrings, and then encoding those patterns using a dictionary of previously seen substrings. The number of distinct sequences found is the Lempel-Ziv index \cite{lempel1976complexity}. In our case we have removed the substrings of length less than 2 as they are uninformative.
Regular signals can be characterized by a small number of patterns and hence have low complexity, while irregular signals are content-rich and therefore less predictable. Lempel-Ziv complexity was introduced to study binary sequences and the ideas introduced were later extended to become the basis of the well-known zip compression algorithm \cite{deutsch1996deflate}.
We have computed compression of a post as the ratio between its Lempel-Ziv complexity index and the total length of the document.
To find the significant structures within a document we have generated an ensemble of 100 documents for each post by randomizing the order of words. We have employed such ensemble as benchmarks to extract the statistically relevant bi-grams for each week by computing the residual occurrence.
We have considered only the statistically relevant bi-grams with respect to the average residual (between 30-40\% of the total bi-grams) and computed the Jaccard similarity index among weeks to assess whether two weeks are statistically similar, i.e. they have the same semantic structures. In this case we have cleaned the text by removing stop-words and punctuation, and considered only the bi-grams with at least 25 occurrences.

Sentiment analysis has been carried out via VADER (\emph{Valence Aware Dictionary and sEntiment Reasoner}) \cite{hutto2014VADER}, a python tool that assigns to each piece of text a score $s$ between -1 (very negative) and +1 (very positive). For each comment/post within a week we have applied VADER to the text and extracted the associated sentiment.
The total emotion of each week has been computed as the total area of the denoised histogram of sentiment, thereby aggregating across bins to potentially account for extreme variations in both positive and negative directions. The denoising of each bin has been carried out by using all the weeks by computing the Z-score, thus revealing weeks with intense sentiment.
\subsection*{Users Analysis}
For each week we have reconstructed the network of social interactions by considering posts and comments. Each user who contributed at least five of these posts/comments during that week is represented as a node, and a direct link between user \emph{i} and \emph{j} is present if \emph{i} commented on posts/comments by \emph{j}. 
User degree is defined as the number of first neighbors (in both directions) in the network.
To frame the changes in the structure of thematic dialogues we focused on dutiful users that interact persistently with more than 10 posts/comments per week and at least in 70\% of the weeks considered. We report the number of users in Supplementary Information 1, Table S4.
To compute the activity frequency of each user, we have considered the ordered sequence of comments and posts of the user. The mean temporal distance between two consecutive contributions by the user gives the activity period, whose inverse defines the activity frequency.
The semantic compression of each user has been computed via the Lempel-Ziv complexity index, as described in the conversations' analysis but on the document obtained by joining all the comments and posts of the user.
To compute the semantic diversity of each user, we have, firstly, trained Word2Vec \cite{mikolov2013distributed} on all the subreddits, using the Python package \emph{gensim} \cite{rehurek2011gensim}. 
Word2Vec has been trained using the continuous bag of words (CBOW) model to learn word embeddings. In this neural network model, the goal is to predict a target word given a set of context words, where the target is the middle word of the context. The context words, represented as one-hot encoding vectors, are fed into an embedding layer, which serves as a lookup table for the corresponding word embeddings (dense vectors). The embeddings are then fed into a shallow neural network to predict the probability distribution over the vocabulary for the target word, and the weights are updated using back-propagation; thus refining the word embeddings of the first input layer (embedding layer).
In this case we have cleaned the text by removing punctuation and stop-words, lowering and stemming it. We have considered an embedding vector of 100 dimensions, with word window 3 and we have ignored all words with total occurrence lower than 4. The total number of words on which the model is trained is approx. 850M and we have trained the neural network till the loss reached a plateau (max 100 epochs).
We have, then, mapped each comment/post to a point in the semantic space, by averaging the embeddings of the words appearing in a given text. The semantic diversity has been computed as
\begin{equation}
	d_u = \sqrt{\frac{1}{N_u}\sum_{i=1}^{N_u}||v_{i,u}-\langle v\rangle_{u}||^2}\,,
\end{equation}
where $v_{i,u}$ is the semantic vector of post/comment $i$ by user $u$ and $\langle v\rangle_u$ is the average semantic vector over the possible $N_u$ posts/comments made by user $u$ during the week considered.


\section*{Declarations}
\begin{itemize}
\item Data and Code Availability:
Reddit conversation data used in this study can be retrieved from the Reddit or Pushshift API at \url{https://www.reddit.com/r/pushshift/} and were retrieved before October 2022.
The code to reproduce the analysis is released on \href{https://github.com/ComplexConnectionsLab/Recurrent_Patterns_Reddit/}{GitHub}.
For inquiries, please contact A.D. \url{antde@dtu.dk}.
\item Acknowledgements
R.D.C. acknowledges Sony CSL Laboratories in Paris for hosting him during part of the research.
G.C. acknowledges support from “Deep ’N Rec” Progetto di Ricerca di Ateneo of University of Rome Tor Vergata. 
\item Author Contributions:
A.D. and A.M. gathered the data.
A.D. performed the analysis.
A.D. and A.M. realised the figures.
G.C. and R.D.C. designed and supervised the analysis.
All the authors discussed the results, wrote the paper and approved the final manuscript.
\item Competing Interests:
The authors declare no competing interests
\end{itemize}
\bibliography{reference}

\pagebreak[2]
\clearpage
\renewcommand{\figurename}{Supplementary Figure}
\renewcommand{\tablename}{Supplementary Table}
\setcounter{figure}{0}
\renewcommand{\thefigure}{S\arabic{figure}}
\renewcommand{\thetable}{S\arabic{table}}

\section*{Supplementary Information, Section 1: Dataset information}\label{sm_dataset_info}
\begin{table}[h!]
    \caption{Large-scale events considered in the analysis with the relative subreddit community.\label{tab_sm_data_events}}%
	\begin{tabular*}{\columnwidth}{@{\extracolsep\fill}llll@{\extracolsep\fill}}
    \toprule
    {\textbf{Subreddit}} & {\textbf{Date}} & {\textbf{Event}} & {\textbf{Label}}\\
    \midrule
       europe & 2020-01-31 &               Brexit & g\\
       europe & 2020-09-10 &     Cyprus Tensions & j \\
       europe & 2020-08-10 &     Belarus Protest & i \\
       europe & 2020-05-01 &       Lockdown Ease & h \\
       europe & 2020-11-03 &              US 2020 & e \\
       europe & 2021-01-06 &        Capitol Hill & f \\
       europe & 2020-03-11 &             COVID-19 & b \\
     politics & 2020-02-05 &         Trump Trial & a\\
     politics & 2020-10-02 &      Trump COVID-19 & d\\
     politics & 2020-11-03 &              US 2020 & e \\
     politics & 2021-01-06 &        Capitol Hill & f \\
     politics & 2020-03-11 &             COVID-19 & b \\
     politics & 2020-06-06 & Black Lives Matter & c\\
          nba & 2020-01-26 &         Kobe Bryant & k \\
          nba & 2020-03-12 &            NBA Stop & l \\
          nba & 2020-12-22 &      Regular Season & p \\
          nba & 2020-07-31 &         NBA Restart & m \\
          nba & 2020-10-11 &               Finals & n \\
          nba & 2020-11-21 &          NBA Trades  & o \\
          nfl & 2020-04-24 &                Draft & u \\
          nfl & 2020-09-11 &        Kickoff Game  & v \\
          nfl & 2020-02-02 &       SuperBowl LIV & s \\
          nfl & 2020-03-18 &          NFL Trades & t \\
          nfl & 2021-01-10 &              PlayOff & w  \\
    \bottomrule
	\end{tabular*}
\end{table}
\begin{table}[h!]
    \caption{Wikipedia pages retrieved and Google Trends keywords queried for the relative subreddit community.\label{tab_sm_data_pages}}%
    \begin{tabular*}{\columnwidth}{@{\extracolsep\fill}lll@{\extracolsep\fill}}
    \toprule
    {\textbf{Subreddit}} & {\textbf{Google Trends Query}} & {\textbf{Wikipedia Pages}}\\
    \midrule
       europe & -- & \url{https://en.wikipedia.org/wiki/2020_in_the_European_Union} \\
     politics & -- & \url{https://en.wikipedia.org/wiki/2020_in_the_United_States} \\
      nba & nba &          \url{https://en.wikipedia.org/wiki/2020–21_NBA_season} \\
      nfl & nfl &              \url{https://en.wikipedia.org/wiki/2020_NFL_season} \\
    \bottomrule
    \end{tabular*}
\end{table}
\begin{table}[h!]
    \caption{Metadata downloaded from Pushshift for each Reddit comment.\label{tab_sm_data_comments}}%
    \begin{tabular*}{\columnwidth}{@{\extracolsep\fill}ll@{\extracolsep\fill}}
		\toprule
		{\textbf{Column}} & {\textbf{Description}}\\
		\midrule
		Author &  Username\\
		Author ID & ID that uniquely identifies each Reddit user\\
		Comment ID & ID that uniquely identifies each comment\\
		Submission ID &	ID of the post under which the comment was made\\
		Parent ID &	ID of the post or ID of the comment to which the given comment is a reply\\
		Text & Text of the comment\\
		UTC & Epoch Unix timestamp of the comment\\
        \bottomrule
    \end{tabular*}
\end{table}
\begin{table}[h!]
    \caption{Number of dutiful users considered in the Users Dynamics analysis, with relative thresholds for the number of comments in a week and the fraction of weeks active.\label{tab_sm_users_ud}}%
    \begin{tabular*}{\columnwidth}{@{\extracolsep\fill}llll@{\extracolsep\fill}}
		\toprule
        {\textbf{Subreddit}} & {\textbf{Users}} & {\textbf{Threshold weeks}} & {\textbf{Threshold Comments}}\\
        \midrule
        politics & 1209 & 0.7 & 10\\
        nba & 416 & 0.7 & 10\\
        nfl & 535 & 0.7 & 10\\
        europe & 403 & 0.5 & 5\\
        \bottomrule
    \end{tabular*}
\end{table}
\pagebreak[2]
\clearpage
\section*{Supplementary Information, Section 2: Daily Z-score to identify highly engaging events}\label{sm_zscore}

\begin{figure*}[h!]
    \centering
    \includegraphics[width=\textwidth]{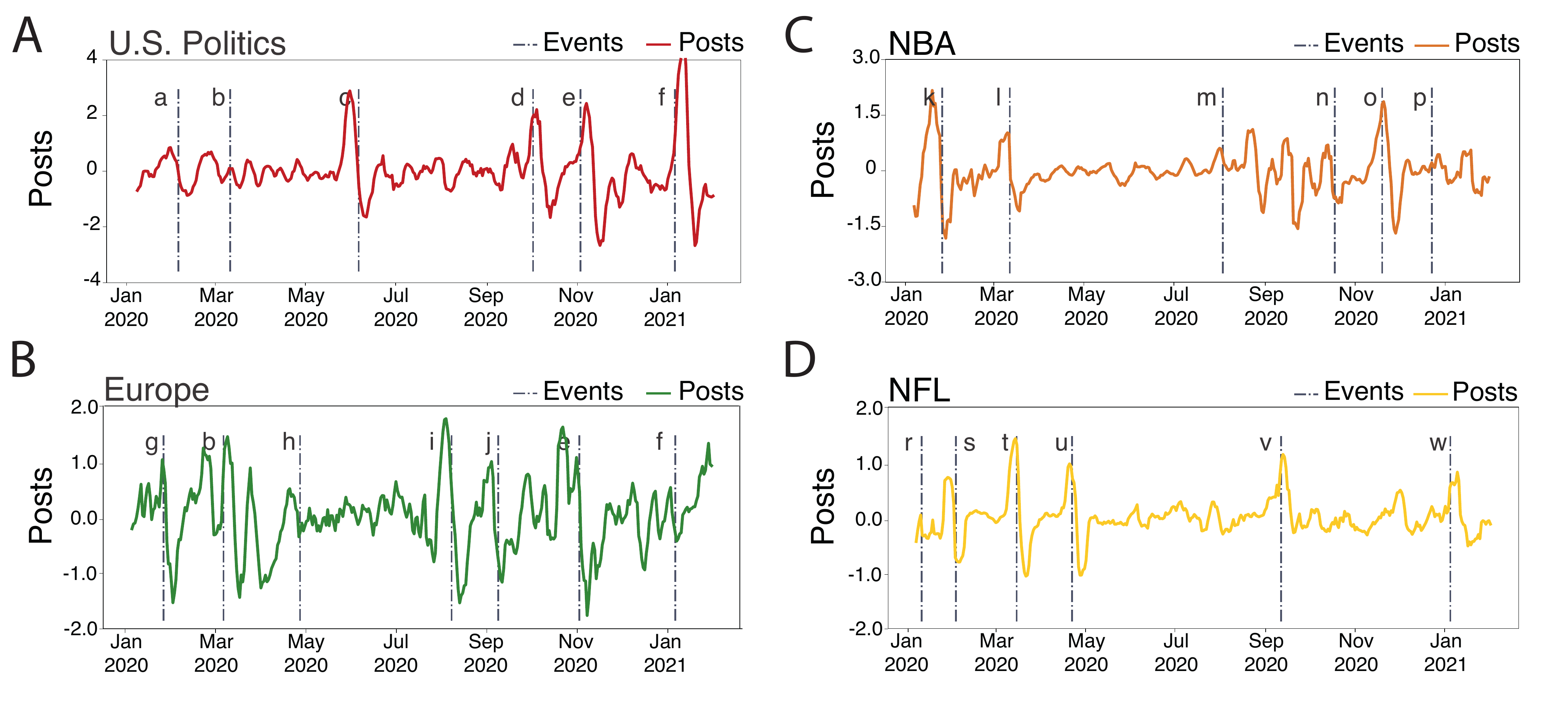}
    \caption{
    Daily Z-score variation of Post Activity in Subreddits. 
    The solid line represents the daily Z-score variation of the number of posts for the following subreddits: U.S. politics (panel A), European (panel B), NBA (panel C), and NFL (panel D). The grey vertical dashed-dotted lines indicate highly engaging events, which correspond to the peaks in the Z-score variation, signifying increased subreddit activity during these times.
    }
    \label{fig_sm_zscore}
\end{figure*}

\pagebreak[2]
\clearpage
\section*{Supplementary Information, Section 3: Null Models of Dynamic Time Warping and Coherence}\label{sm_null_models}
\begin{figure*}[h!]
    \centering
    \includegraphics[width=0.66\textwidth]{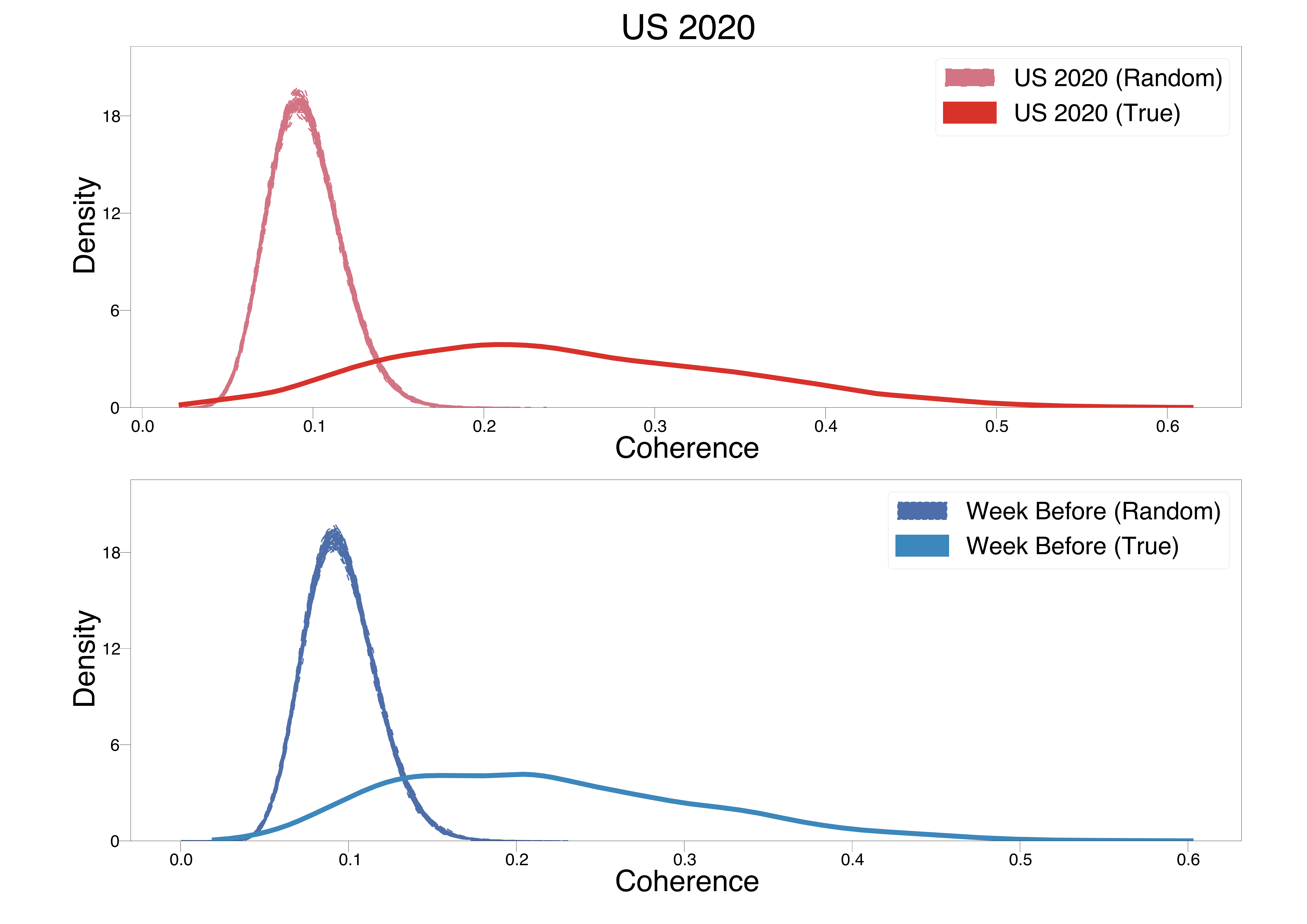}
    \caption{Distributions of coherence distances between the time series of US politics community for the US 2020 election. The dashed lines are the surrogate distributions, while the solid lines are the ground truth. The top panel displays the comparison between the time series of the event week and week before, while the bottom panel displays the case between the week before and two weeks before.}
    \label{fig_sm_null_models_us}
\end{figure*}

\begin{figure*}[h!]
    \centering
    \includegraphics[width=0.66\textwidth]{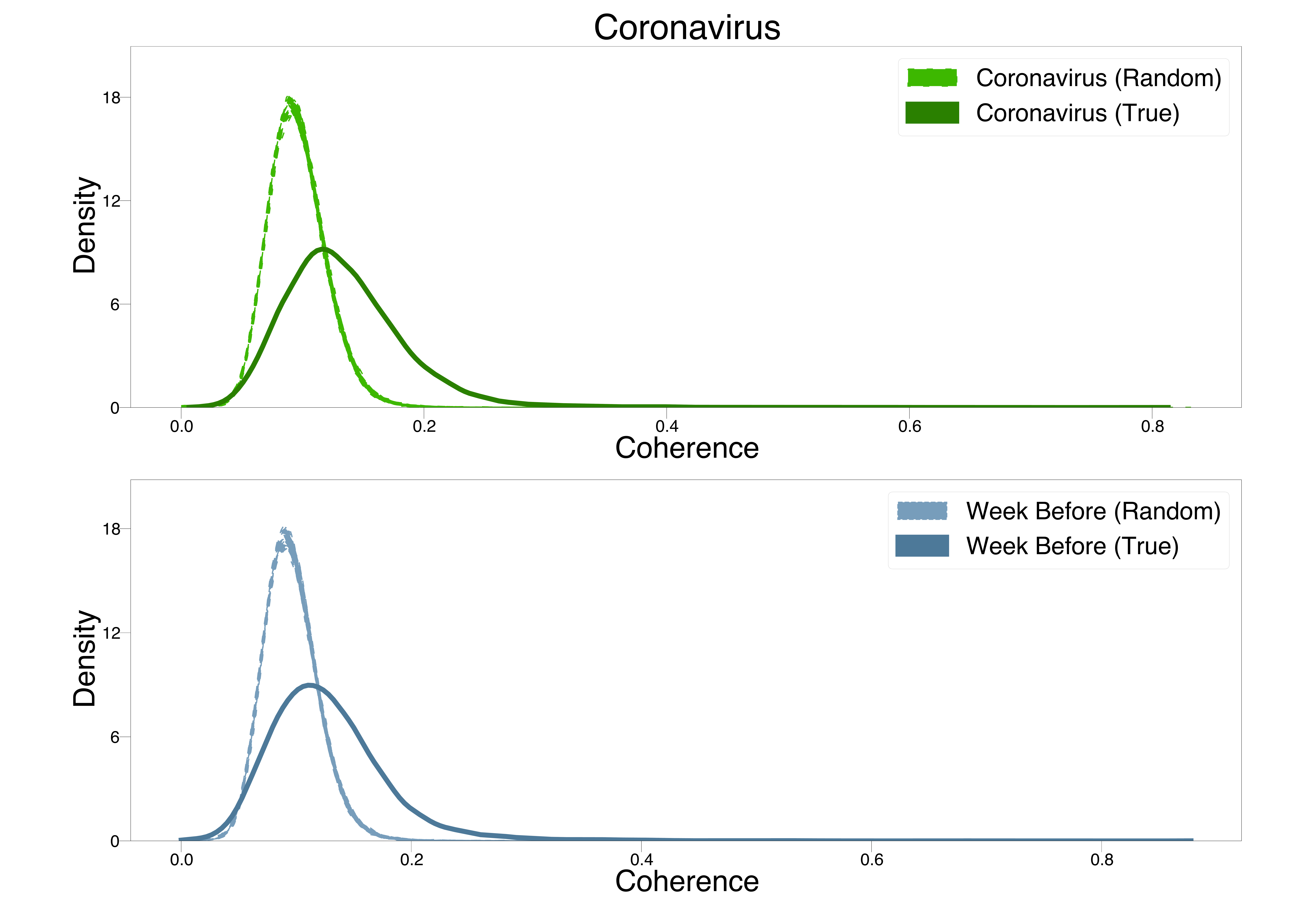}
    \caption{Distributions of coherence distances between the time series of European community for the Coronavirus Outbreak Event. The dashed lines are the surrogate distributions, while the solid lines are the ground truth. The top panel displays the comparison between the time series of the event week and week before, while the bottom panel displays the case between the week before and two weeks before.}
    \label{fig_sm_null_models_eu}
\end{figure*}

\begin{figure*}[h!]
    \centering
    \includegraphics[width=0.66\textwidth]{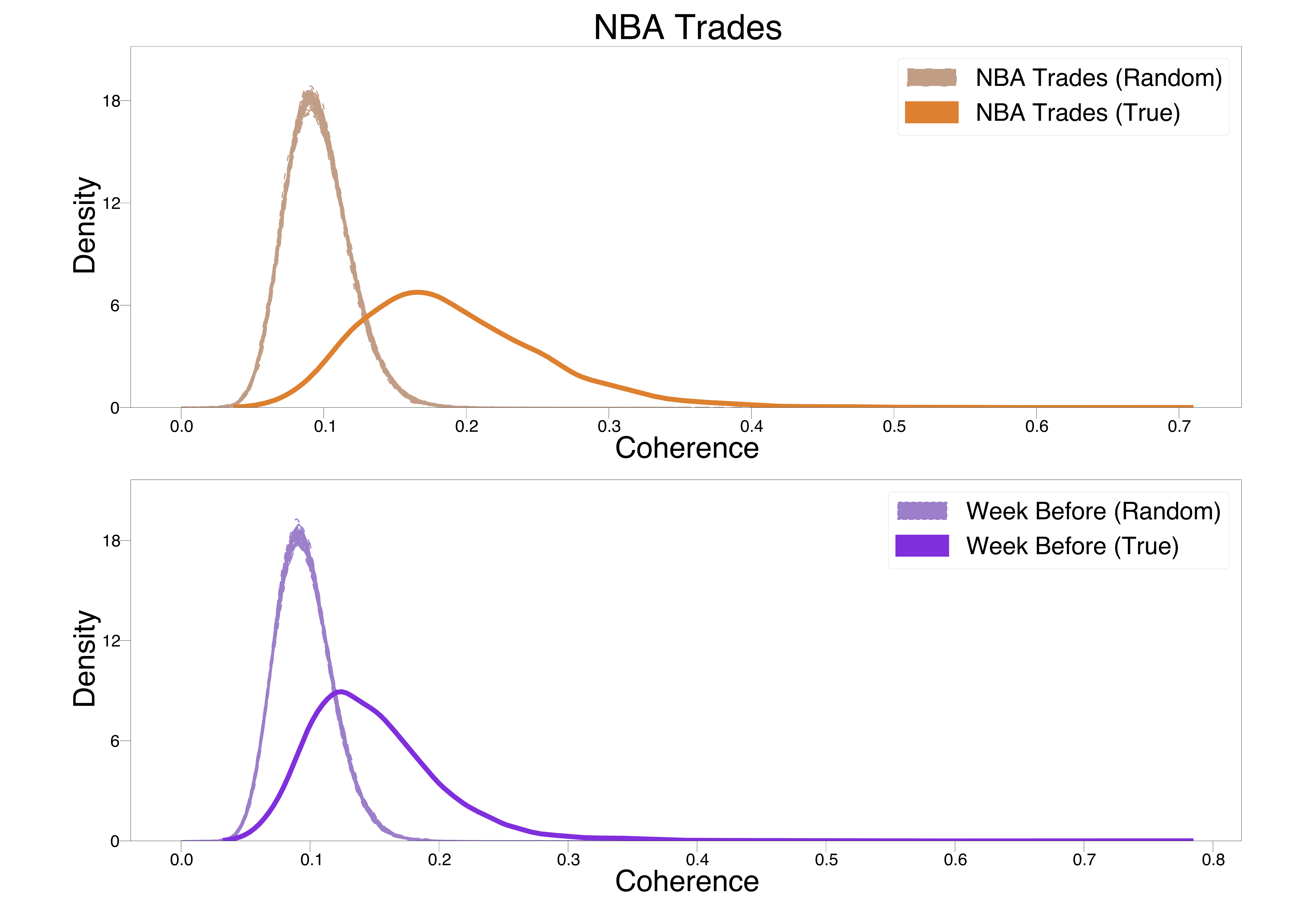}
    \caption{Distributions of coherence distances between the time series of NBA community for the NBA Trades event. The dashed lines are the surrogate distributions, while the solid lines are the ground truth. The top panel displays the comparison between the time series of the event week and week before, while the bottom panel displays the case between the week before and two weeks before.}
    \label{fig_sm_null_models_nba}
\end{figure*}

\begin{figure*}[h!]
    \centering
    \includegraphics[width=0.66\textwidth]{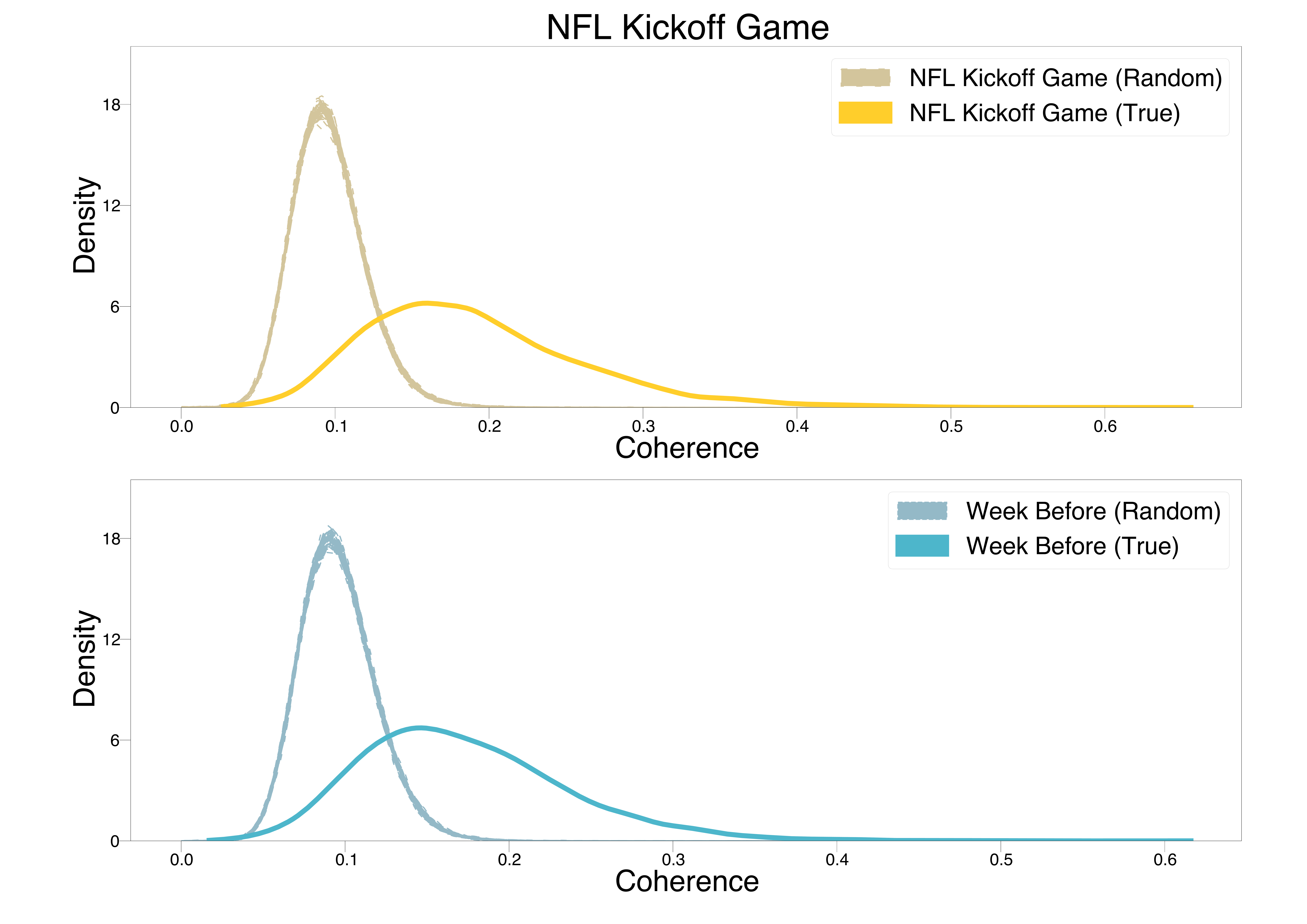}
    \caption{Distributions of coherence distances between the time series of NFL community for the NFL Kickoff game event. The dashed lines are the surrogate distributions, while the solid lines are the ground truth. The top panel displays the comparison between the time series of the event week and week before, while the bottom panel displays the case between the week before and two weeks before.}
    \label{fig_sm_null_models_nfl}
\end{figure*}

To test whether the large variations observed for the Dynamic Time Warping and coherence distances are due to changes of the conversations' temporal structure, we perform a permutation test. We consider for each week the time series $X$ and the surrogate time series $\tilde{X}$, obtained by shuffling the comments' timestamps. For each time series we compute the temporal difference between each comment and its following, then we shuffle these differences and we compute the cumulative sum of the shuffled differences to obtain the surrogate time series. We generate 1000 surrogates for each time series.
As shown in figures \ref{fig_sm_null_models_us}, \ref{fig_sm_null_models_eu}, \ref{fig_sm_null_models_nba}, \ref{fig_sm_null_models_nfl}, the distributions of the coherence distances between the surrogate time series are different from the real ones: they display a sharper peak around 0.1.
\newpage
\pagebreak[2]
\clearpage
\section*{Supplementary Information, Section 4: Time Metrics}\label{sm_time_metrics}
The distributions of the reply speeds become sharper during large-scale events as displayed by the solid lines in Supplementary Figures \ref{fig_sm_time_metrics_distrib_dt_1},\ref{fig_sm_time_metrics_distrib_dt_2},\ref{fig_sm_time_metrics_distrib_dt_3},\ref{fig_sm_time_metrics_distrib_dt_4}. The fit of the distributions is not the scope of this work, however we find that data are well approximated by log-normal distributions. We report for several weeks and subreddits the residual sum of squares (RSS) in Table 5.
As shown in the panels of Supplementary Figure \ref{fig_sm_time_metrics_average}, the standard deviation of the reply speed significantly decreases during exogenous events.

\begin{table}[h!]
    \caption{Residual Sum of Squares (RSS) for the reply speeds for different distributions (Gamma, Log-Normal, Powerl-Law) for several weeks for all the subreddit.\label{tab_sm_users}}%
    \begin{tabular*}{\columnwidth}{@{\extracolsep\fill}llll@{\extracolsep\fill}}
		\toprule
        {\textbf{Subreddit}} & {\textbf{Date}}&{\textbf{Distribution}} & {\textbf{RSS}}\\
        \midrule
        nba & 2020-11-21 & Log-Normal & 0.04\\
        nba & 2020-11-21 & Power-Law & 0.37\\
        nba & 2020-11-21 & Gamma & 0.18\\
        nba & 2020-03-12 & Log-Normal & 0.09\\
        nba & 2020-03-12 & Power-Law & 0.92\\
        nba & 2020-03-12 & Gamma & 0.48\\
        nfl & 2020-09-11 & Log-Normal & 0.42\\
        nfl & 2020-09-11 & Power-Law & 2.65\\
        nfl & 2020-09-11 & Gamma & 1.81\\
        nfl & 2020-04-24 & Log-Normal & 0.20\\
        nfl & 2020-04-24 & Power-Law & 1.54\\
        nfl & 2020-04-24 & Gamma & 1.02\\
        europe & 2020-01-31 & Log-Normal & 0.03\\
        europe & 2020-01-31 & Power-Law & 0.19\\
        europe & 2020-01-31 & Gamma & 0.1\\
        europe & 2020-08-10 & Log-Normal & 0.03\\
        europe & 2020-08-10 & Power-Law & 0.3\\
        europe & 2020-08-10 & Gamma & 0.15\\
        politics & 2020-06-06 & Log-Normal & 0.26\\
        politics & 2020-06-06 & Power-Law & 0.54\\
        politics & 2020-06-06 & Gamma & 0.33\\
        politics & 2020-11-03 & Log-Normal & 0.52\\
        politics & 2020-11-03 & Power-Law & 0.65\\
        politics & 2020-11-03 & Gamma & 0.8\\
        \bottomrule
    \end{tabular*}
\end{table}

\begin{figure*}[h!]
    \centering
    \includegraphics[width=0.66\textwidth]{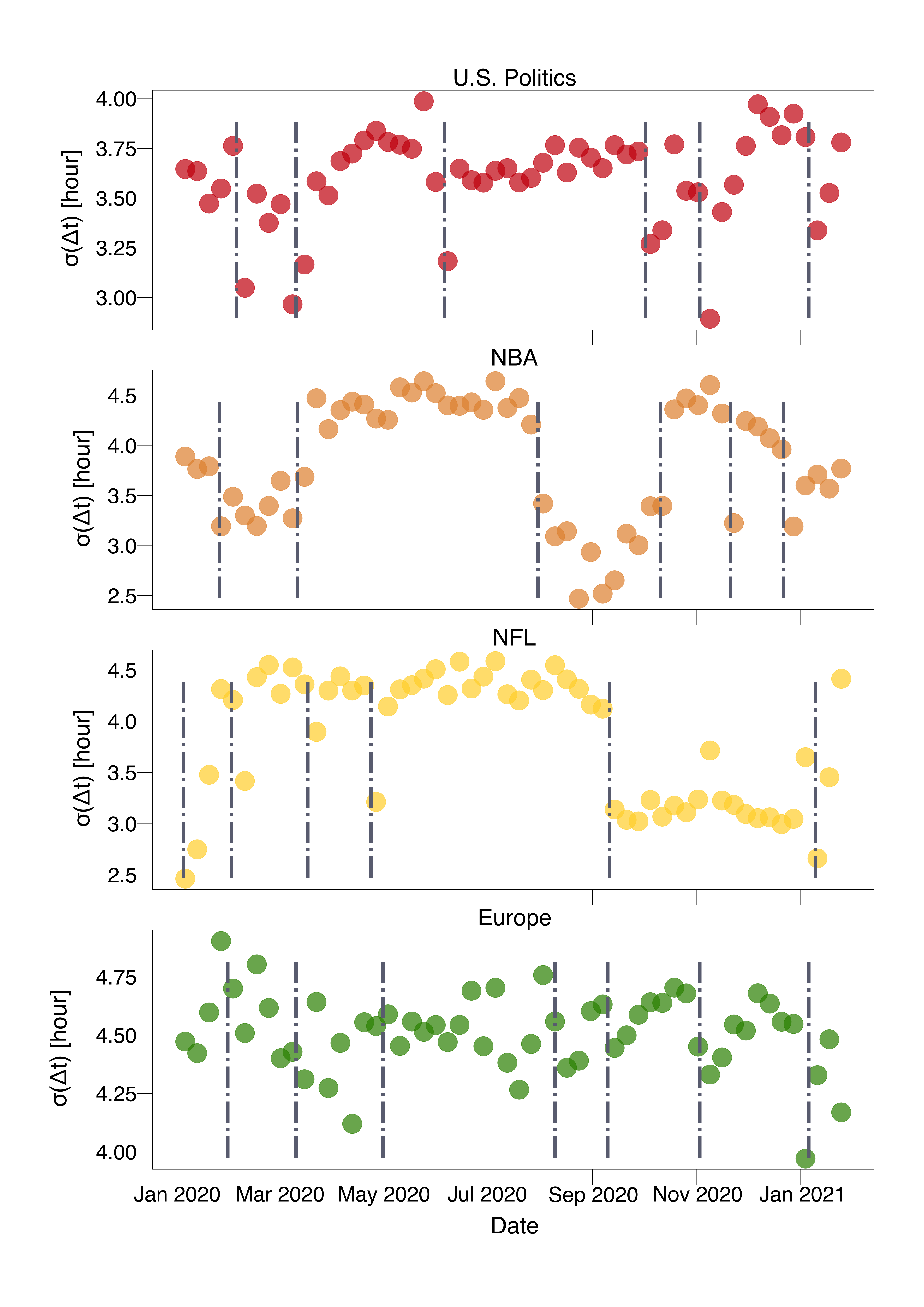}
    \caption{Standard deviation of the reply speed of each week for the analysed communities (Panels). The vertical grey lines mark the large-scale events. }
    \label{fig_sm_time_metrics_average}
\end{figure*}
\newpage
\begin{figure*}[h!]
    \centering
    \includegraphics[width=0.66\textwidth]{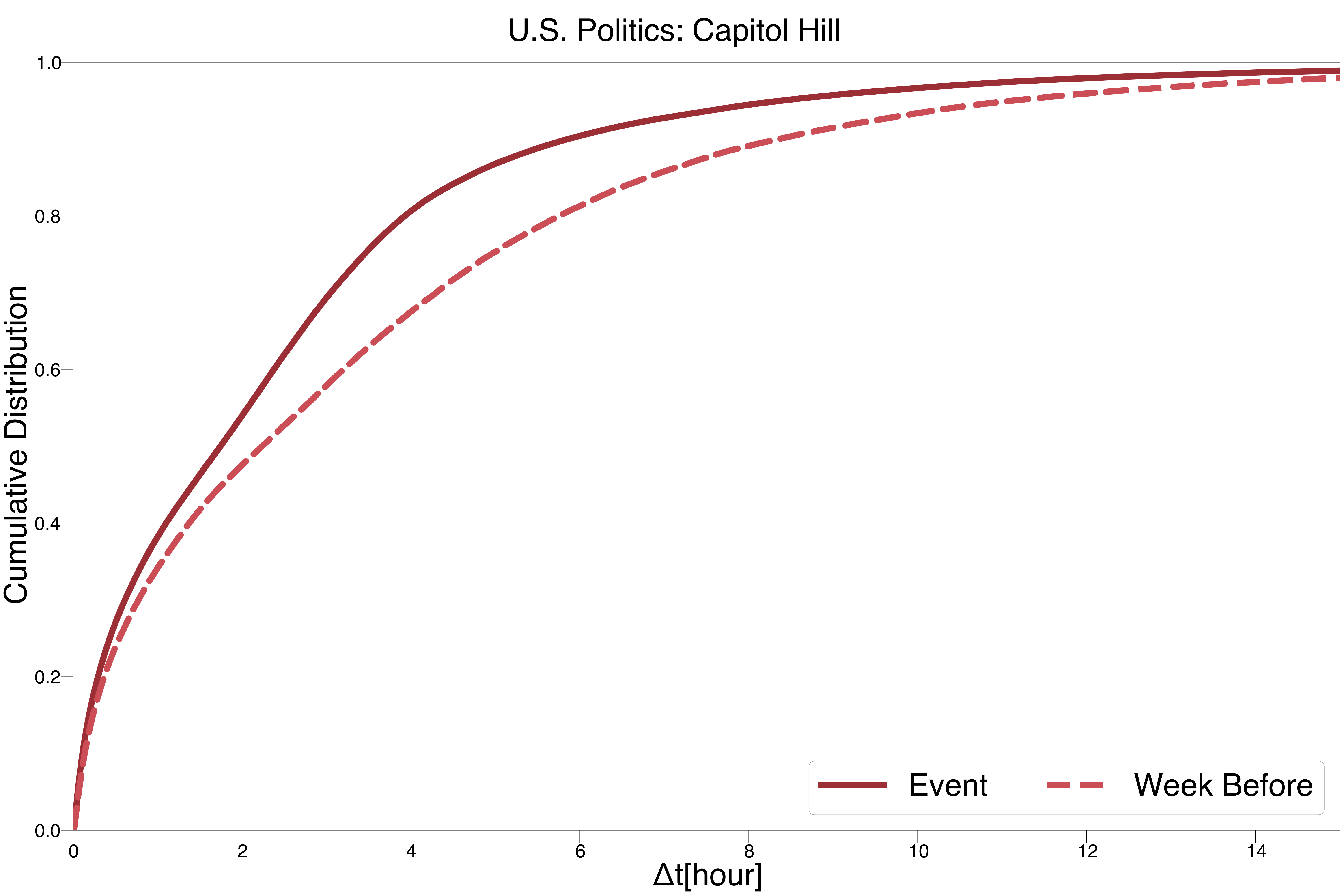}
    \caption{Cumulative distributions of the answering times for the US politics community during the Capitol Hill event (solid line) and the week before (dashed line).}
    \label{fig_sm_time_metrics_distrib_dt_1}
\end{figure*}
\begin{figure*}[h!]
    \centering
    \includegraphics[width=0.66\textwidth]{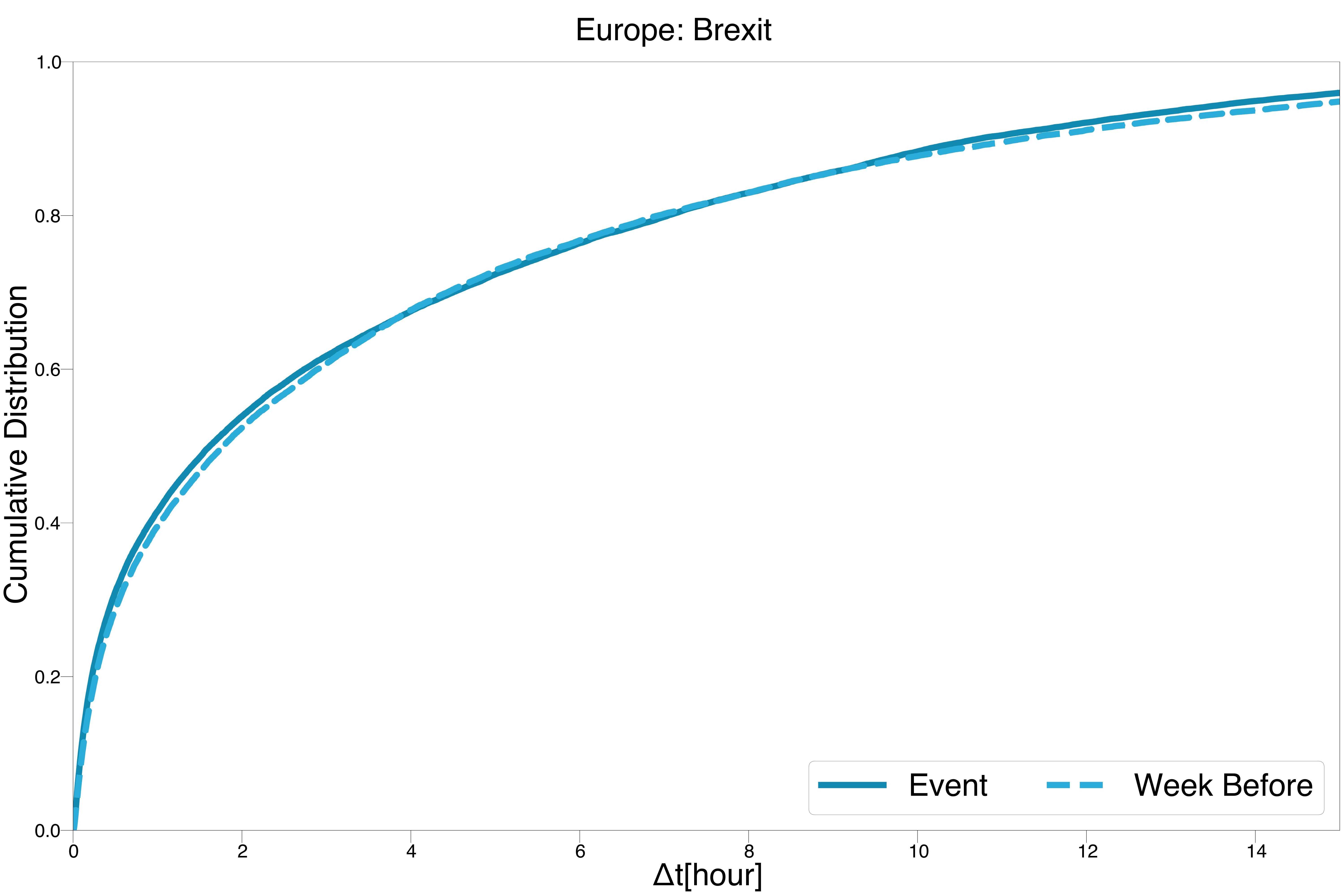}
    \caption{Cumulative distributions of the answering times for the European community during the Brexit event (solid line) and the week before (dashed line).}
    \label{fig_sm_time_metrics_distrib_dt_2}
\end{figure*}
\newpage
\begin{figure*}[h!]
    \centering
    \includegraphics[width=0.66\textwidth]{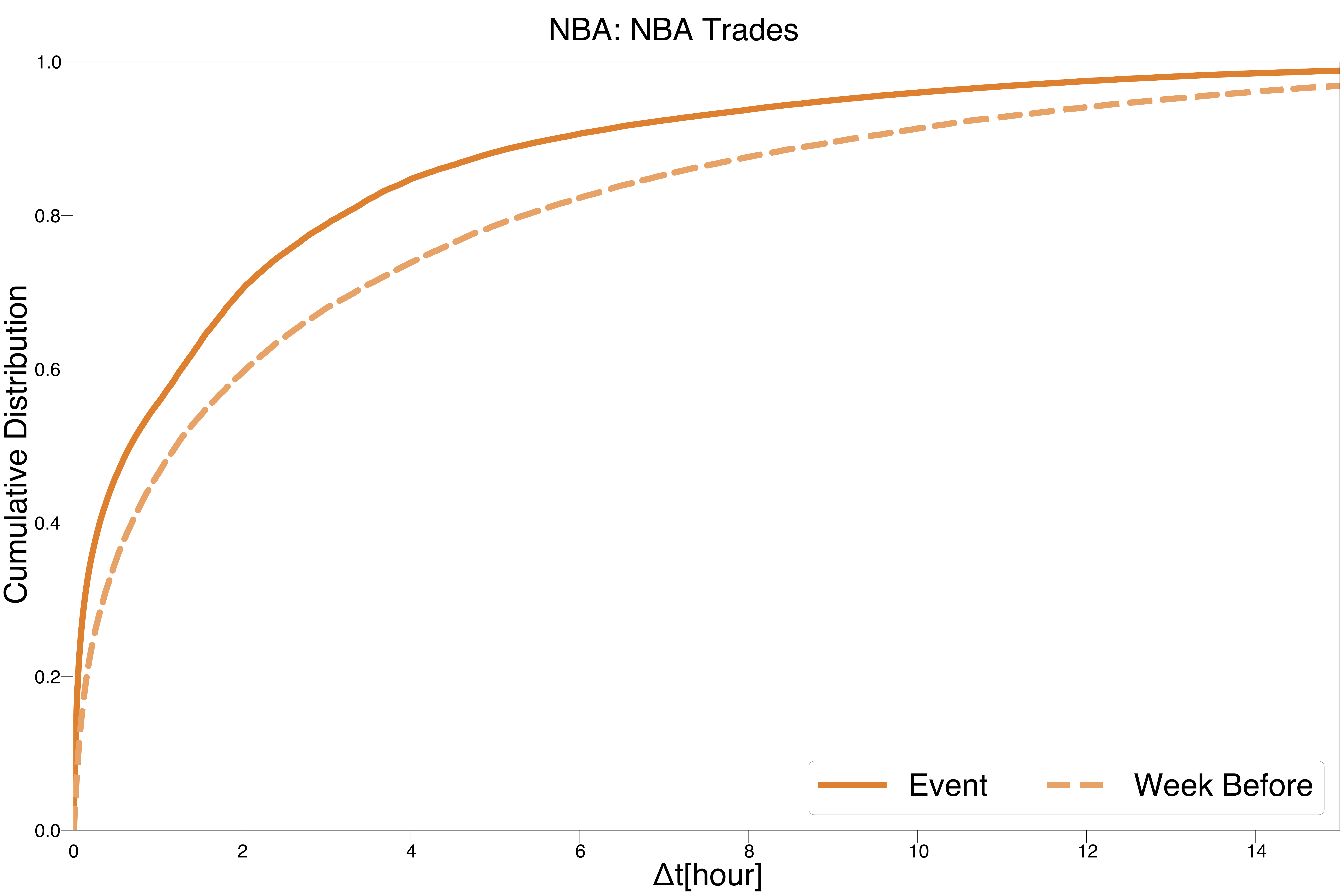}
    \caption{Cumulative distributions of the answering times for the NBA community during the NBA trades event (solid line) and the week before (dashed line).}
    \label{fig_sm_time_metrics_distrib_dt_3}
\end{figure*}
\begin{figure*}[h!]
    \centering
    \includegraphics[width=0.66\textwidth]{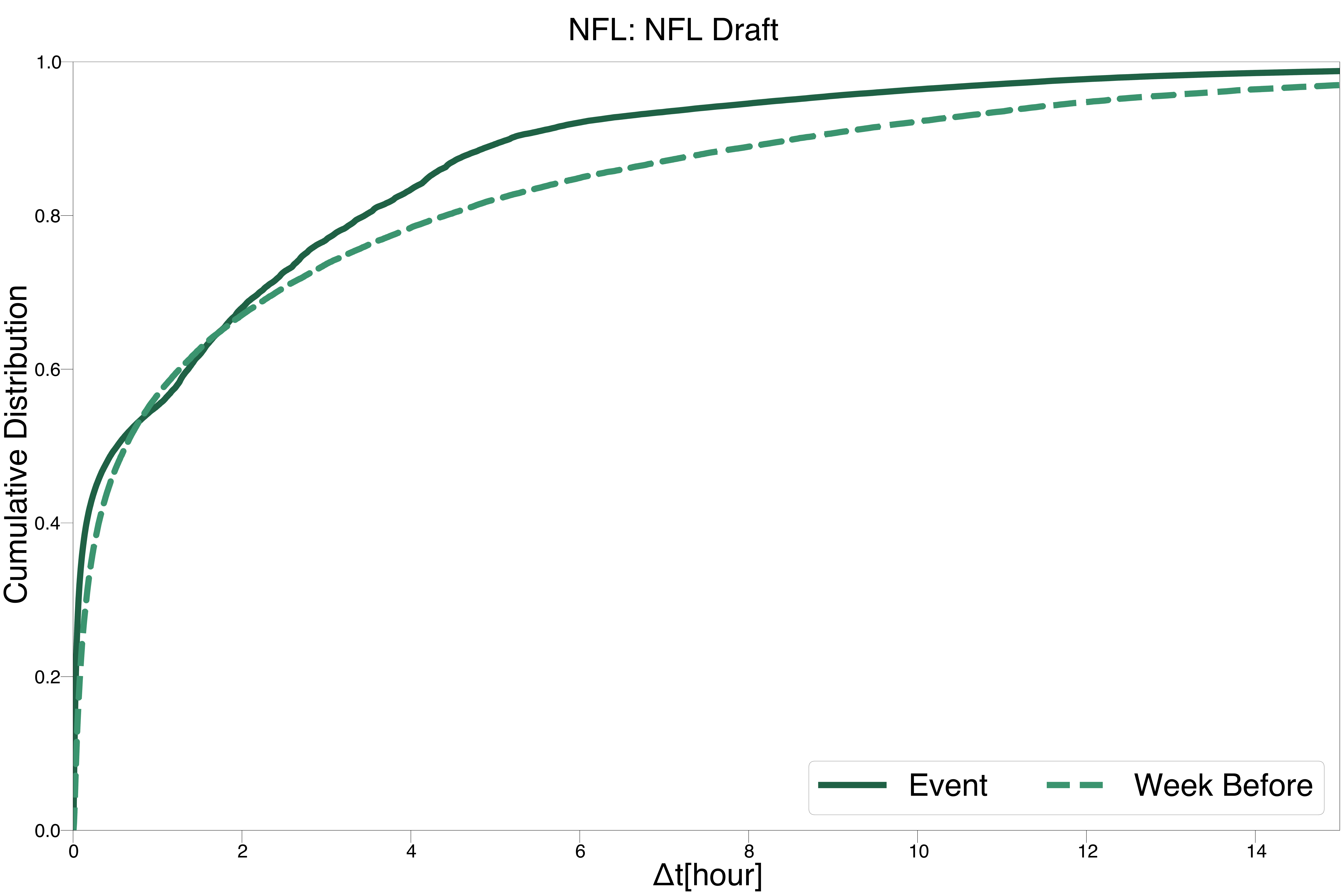}
    \caption{Cumulative distributions of the answering times for the NFL community during the NFL draft event (solid line) and the week before (dashed line).}
    \label{fig_sm_time_metrics_distrib_dt_4}
\end{figure*}
\newpage
\begin{figure*}[h!]
    \centering
    \includegraphics[width=0.66\textwidth]{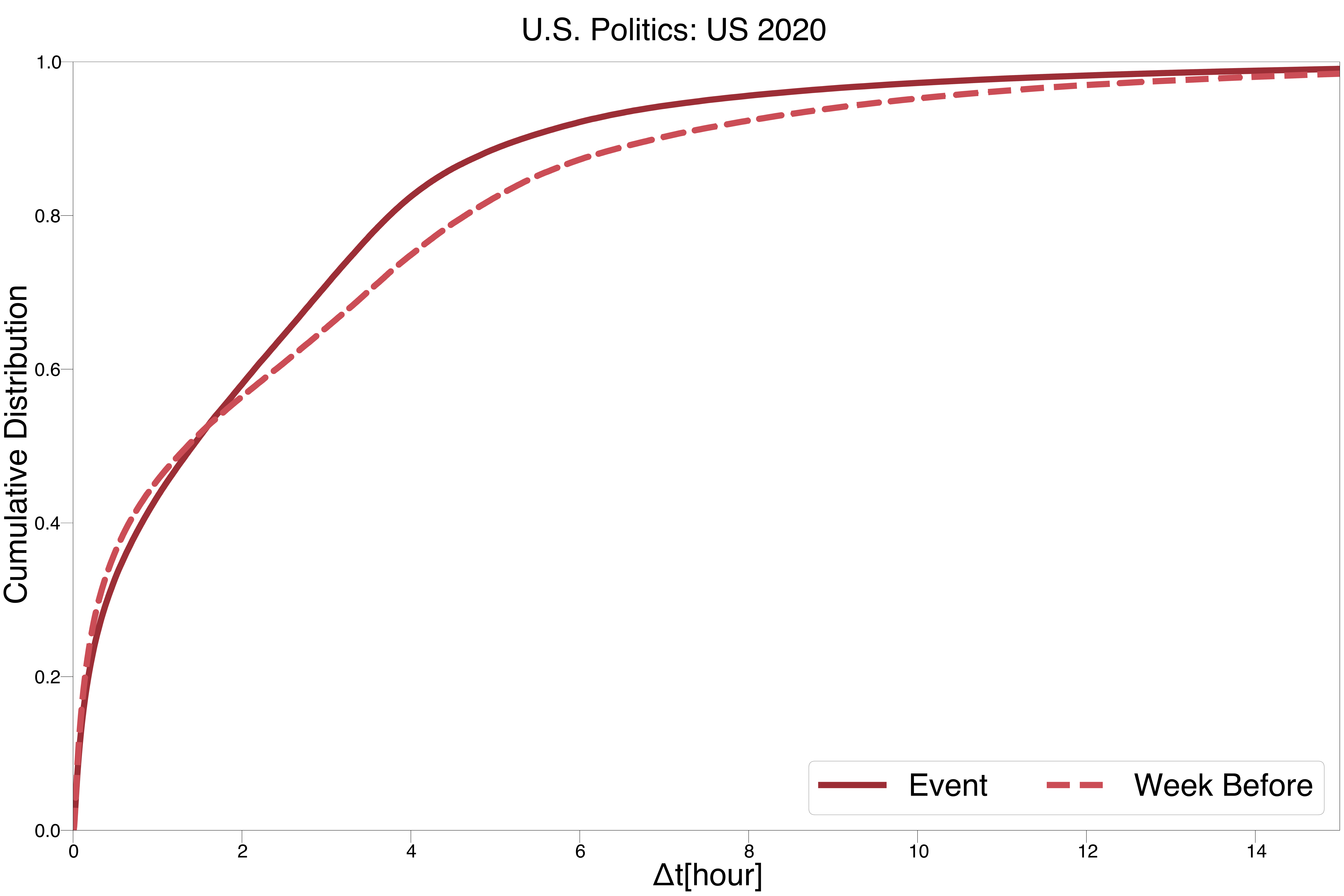}
    \caption{Cumulative distributions of the answering times for the US politics community during the US 2020 election event (solid line) and the week before (dashed line).}
    \label{fig_sm_time_metrics_distrib_dt_5}
\end{figure*}
\begin{figure*}[h!]
    \centering
    \includegraphics[width=0.66\textwidth]{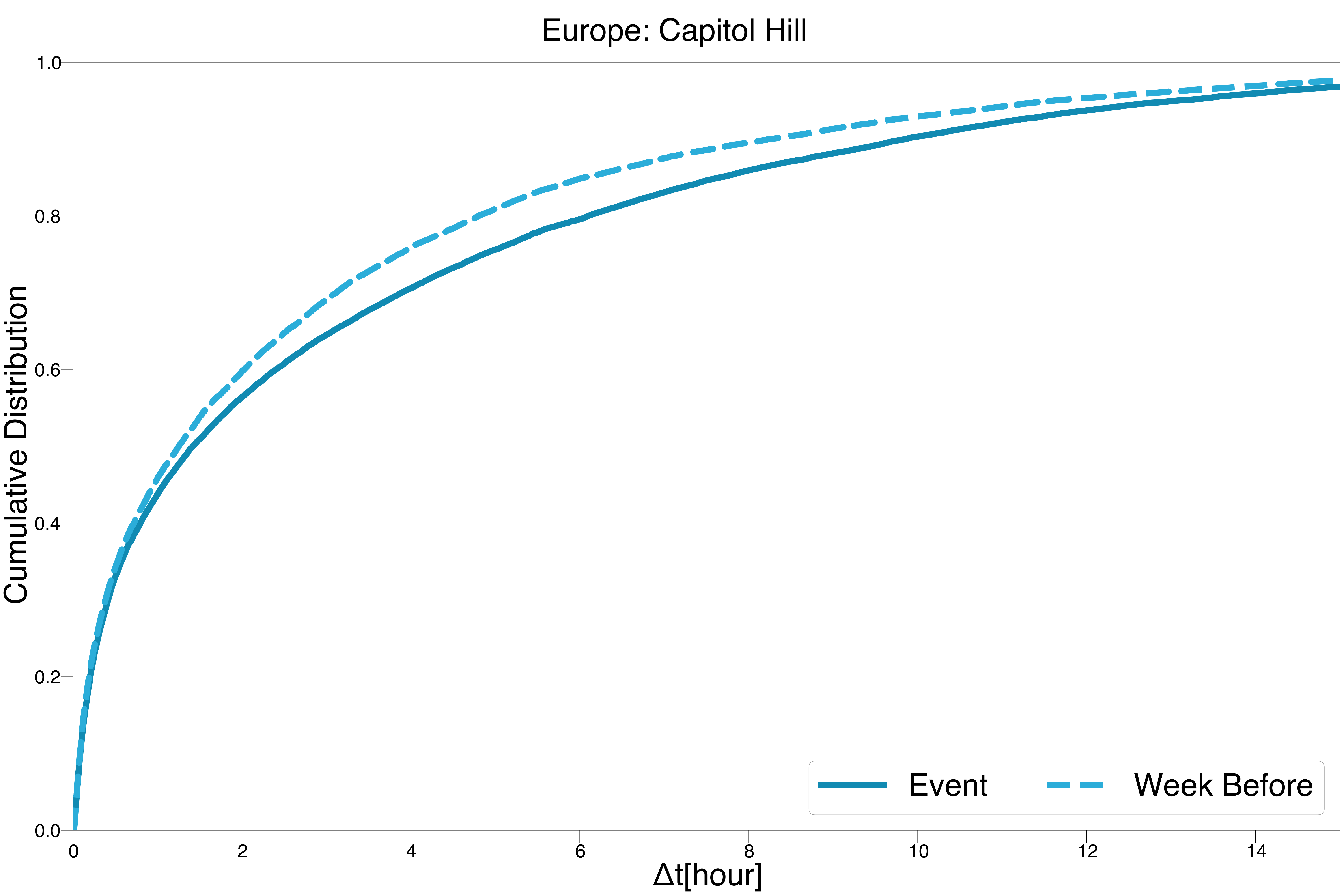}
    \caption{Cumulative distributions of the answering times for the European community during the Capitol Hill event (solid line) and the week before (dashed line).}
    \label{fig_sm_time_metrics_distrib_dt_6}
\end{figure*}
\pagebreak[2]
\clearpage
\section*{Supplementary Information, Section 5: Dissimilarity index between weeks}\label{sm_ts_jaccard}
\begin{figure*}[h!]
    \centering
    \includegraphics[width=\textwidth]{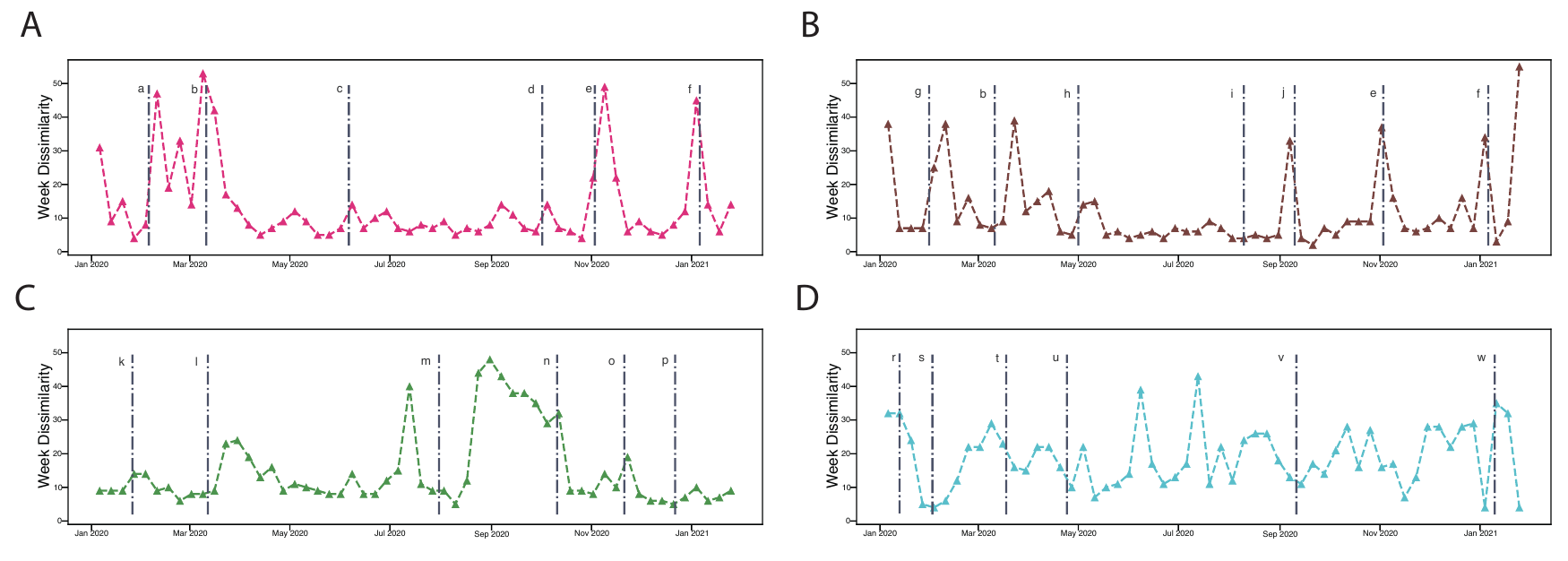}
    \caption{
    Dissimilarity indices derived from Jaccard index.
    The dashed lines represent the weekly dissimilarity indices for the following subreddits: U.S. politics (panel A), European (panel B), NBA (panel C), and NFL (panel D).
    This figure illustrates the weekly dissimilarity indices, which represent the number of weeks where the Jaccard index falls below the median value. Distinctive peaks in the graph correspond to weeks with high engagement (grey vertical dashed-dotted lines) and unique bi-gram usage, indicating the influence of significant events or topics on language patterns.    
    }
    \label{fig_sm_jaccard}
\end{figure*}
\pagebreak[2]
\clearpage
\section*{Supplementary Information, Section 6: Users dynamics for other events}\label{sm_users_dynamics}
We report the users' changes of the variables explained in the main text (semantic diversity, compression, activity frequency and degree) for several exogenous events and different communities.
\begin{figure*}[h!]
    \centering
    \includegraphics[width=0.66\textwidth]{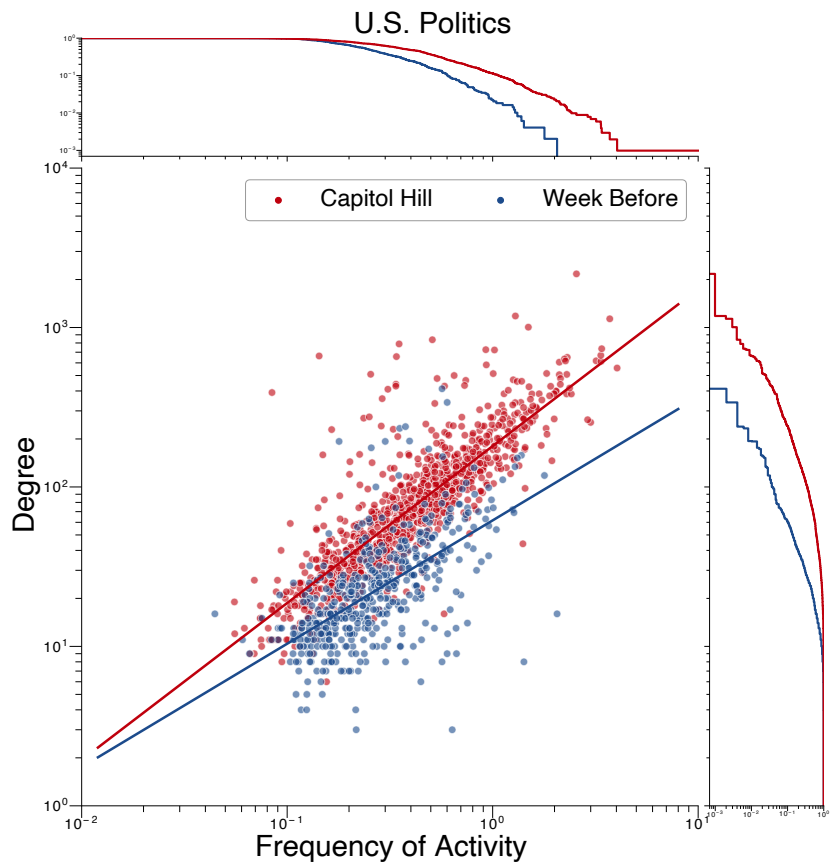}
    \caption{The event considered is the Capitol Hill event for the American community. In the central panels it is shown the relation between the frequency of activity of each user and the interacting peers (degree). In the marginal plots it is reported the survival function of each variable and each week.}
    \label{fig_sm_users_dynamics_fd_us}
\end{figure*}
\begin{figure*}[h!]
    \centering
    \includegraphics[width=0.66\textwidth]{supplementary/users_time_degree/users_freq_degree_nfl_kickoff_game.pdf}
    \caption{The event considered is the NFL Kickoff Game event for the NFL community. In the central panels it is shown the relation between the frequency of activity of each user and the interacting peers (degree). In the marginal plots it is reported the survival function of each variable and each week.}
    \label{fig_sm_users_dynamics_fd_nfl}
\end{figure*}
\newpage
\begin{figure*}[h!]
    \centering
    \includegraphics[width=0.66\textwidth]{supplementary/users_time_degree/users_freq_degree_orlando.pdf}
    \caption{The event considered is the Orlando event for the NBA community. In the central panels it is shown the relation between the frequency of activity of each user and the interacting peers (degree). In the marginal plots it is reported the survival function of each variable and each week.}
    \label{fig_sm_users_dynamics_fd_nba}
\end{figure*}
\begin{figure*}[h!]
    \centering
    \includegraphics[width=0.66\textwidth]{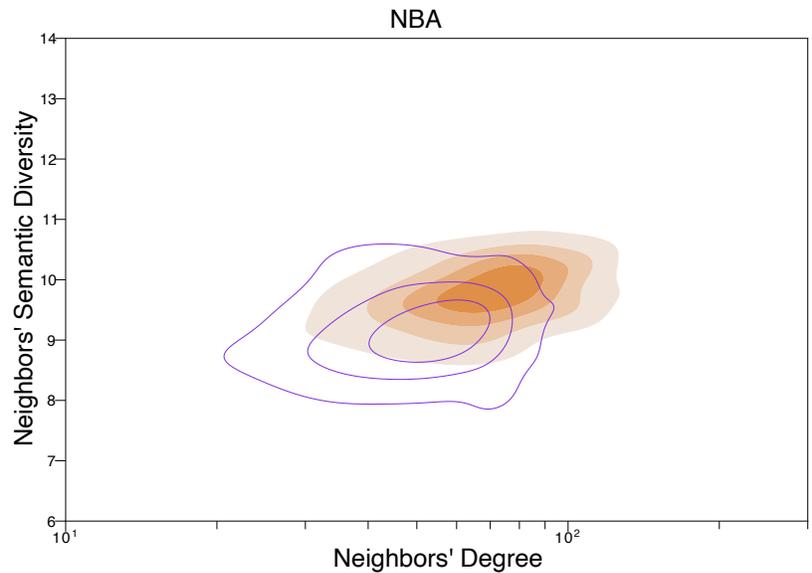}
    \caption{The event considered is the Orlando event (Restart NBA) for the NBA community. The density plots show the variations of the peers' degree and semantic diversity.}
    \label{fig_sm_users_dynamics_nd_nba}
\end{figure*}
\newpage
\begin{figure*}[h!]
    \centering
    \includegraphics[width=0.66\textwidth]{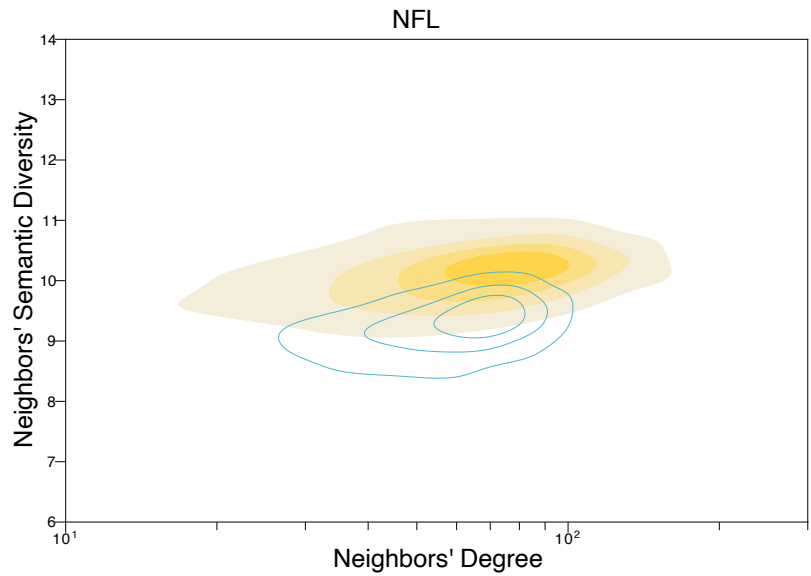}
    \caption{The event considered is the NFL Kickoff Game event for the NFL community. The density plots show the variations of the peers' degree and semantic diversity.}
    \label{fig_sm_users_dynamics_nd_nfl}
\end{figure*}
\begin{figure*}[h!]
    \centering
    \includegraphics[width=0.66\textwidth]{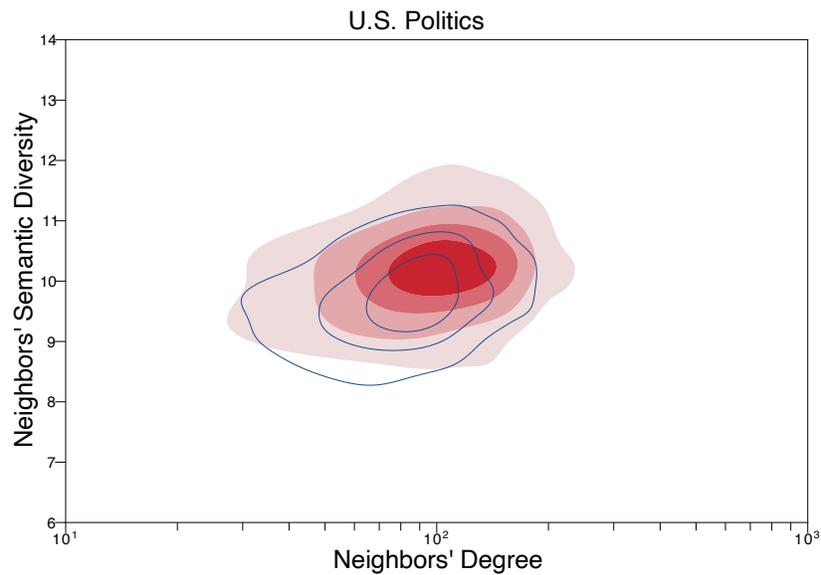}
    \caption{The event considered is the Trump Trial event for the American community. The density plots show the variations of the peers' degree and semantic diversity.}
    \label{fig_sm_users_dynamics_nd_us}
\end{figure*}
\newpage
\begin{figure*}[h!]
    \centering
    \includegraphics[width=0.66\textwidth]{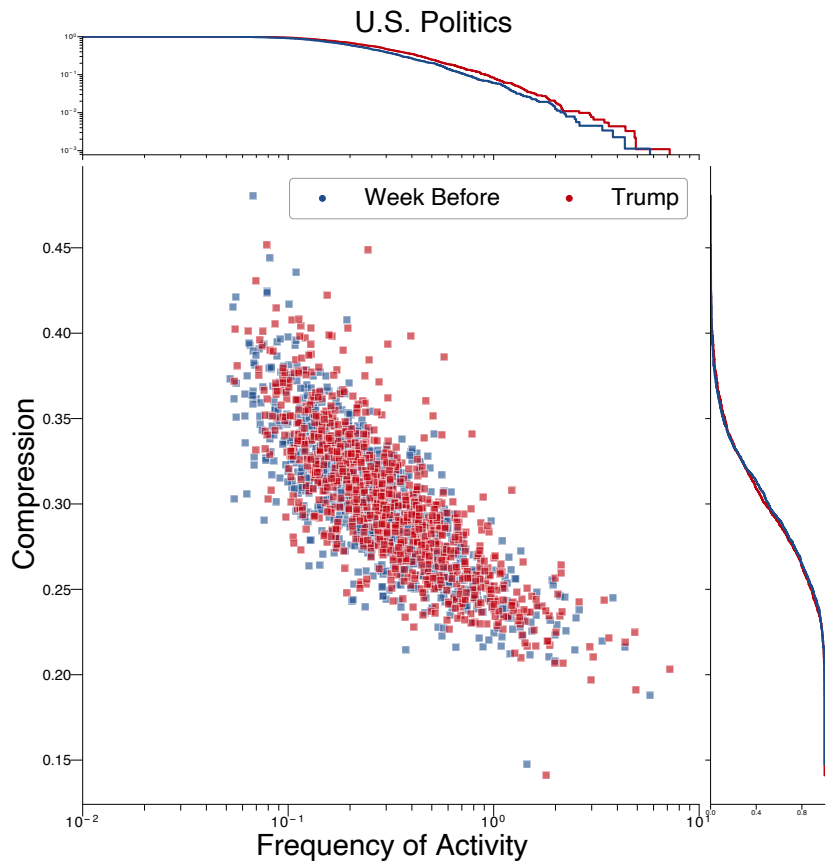}
    \caption{The event considered is the Trump Trial event for the American community. In the central panels it is shown the relation between the compression and the frequency of activity. In the marginal plots it is reported the survival function of each variable and each week.}
    \label{fig_sm_users_dynamics_fc_us}
\end{figure*}
\begin{figure*}[h!]
    \centering
    \includegraphics[width=0.66\textwidth]{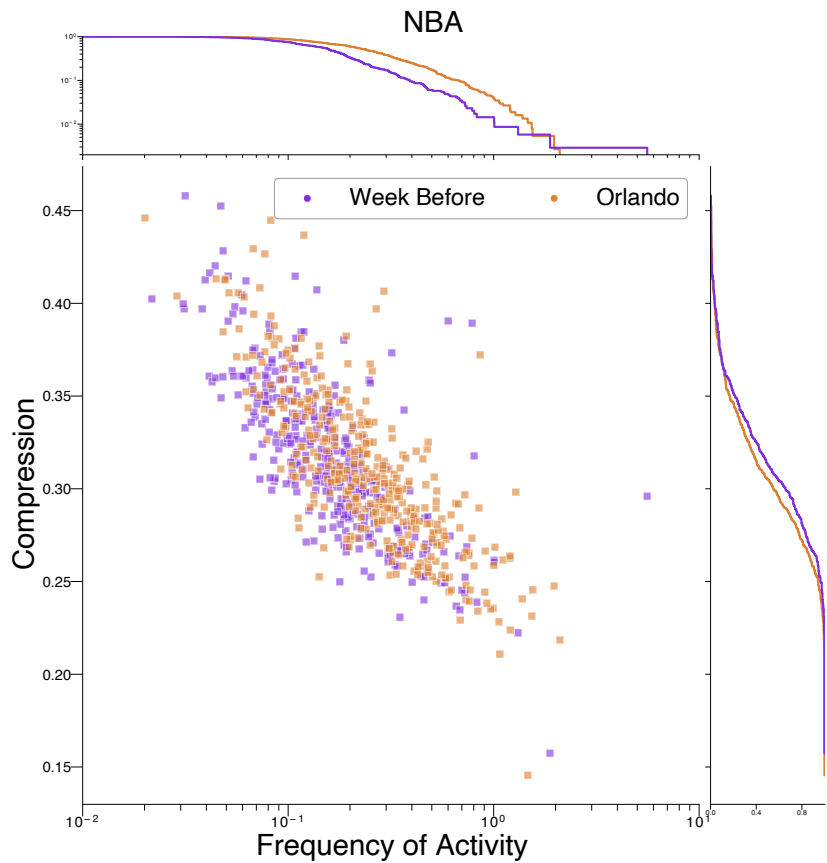}
    \caption{The event considered is the Orlando event (Restart NBA) for the NBA community. In the central panels it is shown the relation between the compression and the frequency of activity. In the marginal plots it is reported the survival function of each variable and each week.}
    \label{fig_sm_users_dynamics_fc_nba}
\end{figure*}
\newpage
\begin{figure*}[h!]
    \centering
    \includegraphics[width=0.66\textwidth]{supplementary/users_compression/users_freq_compression_nfl_kickoff_game.pdf}
    \caption{The event considered is the NFL Kickoff Game event for the NFL community. In the central panels it is shown the relation between the compression and the frequency of activity. In the marginal plots it is reported the survival function of each variable and each week.}
    \label{fig_sm_users_dynamics_fc_nfl}
\end{figure*}

\newpage
\begin{figure*}[h!]
    \centering
    \includegraphics[width=0.75\textwidth]{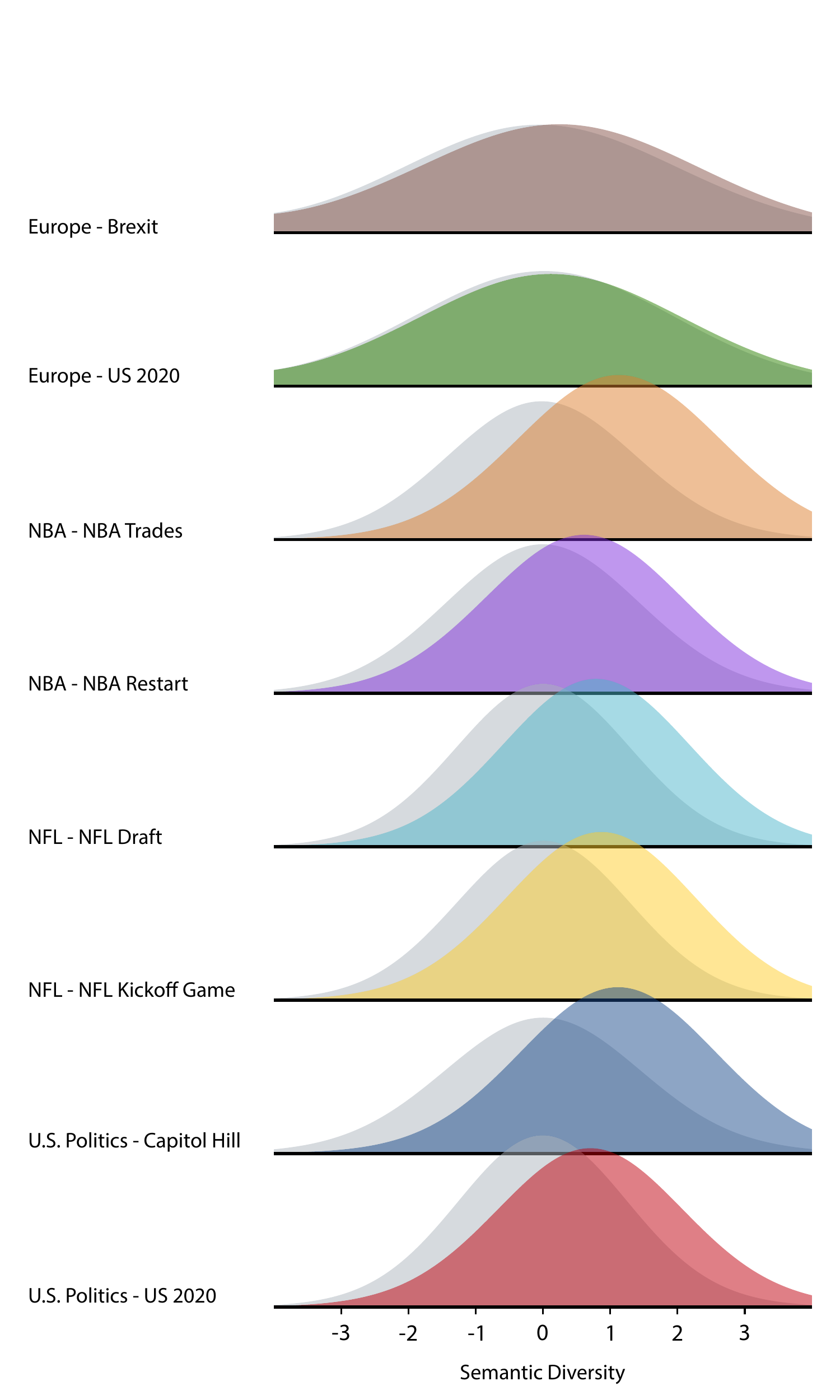}
    \caption{
    Distribution of semantic diversity across various community-event combinations.
    Each panel represents a specific combination, with the week prior to the event shaded in grey. Data has been z-scored relative to the preceding week to highlight deviations in semantic diversity.
    }
    \label{fig_sm_joyplot}
\end{figure*}

\begin{table*}[ht]
\centering
\caption{
Coefficients of Power-law Regression $\lambda$ and Semantic Diversity Metrics for Events Across Subreddits. We report the regression coefficients (with 95\% confidence intervals) for each event and subreddit during the week of the event (Event) and the week before (Week Before). We report the median and interquartile range (IQR) of semantic diversity scores for both time periods.
}
\label{tab_main}
\resizebox{\textwidth}{!}{ 
\begin{tabular}{|l|l|c|c|c|c|c|c|c|c|}
\hline
Event & Subreddit & $\lambda$ & 95\% CI & $\lambda$ & 95\% CI & Median & IQR & Median & IQR \\ 
 &  & \multicolumn{2}{c|}{Event} & \multicolumn{2}{c|}{Week Before} & \multicolumn{2}{c|}{Event} & \multicolumn{2}{c|}{Week Before} \\
\hline
Kobe Bryant & NBA & 0.99 & [0.94, 1.05] & 0.90 & [0.83, 0.98] & 9.86 & 0.67 & 9.57 & 0.75 \\
NBA Stop & NBA & 0.91 & [0.84, 0.99] & 0.94 & [0.87, 1.0] & 9.97 & 0.75 & 9.86 & 0.82 \\
Regular Season & NBA & 0.95 & [0.89, 1.01] & 0.93 & [0.87, 0.99] & 10.12 & 0.79 & 9.59 & 0.77 \\
NBA Restart & NBA & 0.88 & [0.82, 0.94] & 0.73 & [0.65, 0.81] & 9.81 & 0.73 & 9.19 & 0.86 \\
NBA Finals & NBA & 0.80 & [0.73, 0.87] & 0.99 & [0.93, 1.05] & 10.02 & 0.90 & 9.89 & 0.75 \\
NBA Trades & NBA & 1.04 & [0.99, 1.09] & 0.79 & [0.71, 0.86] & 10.04 & 0.62 & 8.97 & 0.75 \\
Brexit & Europe & 0.65 & [0.55, 0.75] & 0.53 & [0.42, 0.64] & 8.9 & 1.11 & 8.69 & 1.1 \\
Cyprus Tensions & Europe & 0.59 & [0.49, 0.69] & 0.52 & [0.43, 0.62] & 8.48 & 1.07 & 8.66 & 1.06 \\
Belarus Protest & Europe & 0.62 & [0.52, 0.71] & 0.69 & [0.59, 0.79] & 8.67 & 0.89 & 8.61 & 1.15 \\
Lockdown Ease & Europe & 0.47 & [0.38, 0.57] & 0.55 & [0.45, 0.65] & 8.47 & 1.15 & 8.53 & 0.95 \\
US 2020 & Europe & 0.66 & [0.56, 0.76] & 0.66 & [0.56, 0.75] & 9.04 & 0.93 & 8.94 & 1.21 \\
Capitol Hill & Europe & 0.49 & [0.35, 0.63] & 0.33 & [0.22, 0.45] & 8.21 & 1.02 & 8.89 & 1.26 \\
COVID-19 & Europe & 0.64 & [0.56, 0.72] & 0.66 & [0.57, 0.74] & 8.99 & 1.11 & 9.01 & 1.18 \\
Trump Trial & U.S. Politics & 0.96 & [0.92, 0.99] & 0.94 & [0.9, 0.98] & 10.19 & 1.04 & 9.82 & 0.97 \\
Trump COVID-19 & U.S. Politics & 1.00 & [0.96, 1.05] & 0.93 & [0.88, 0.97] & 10.02 & 0.82 & 9.76 & 0.86 \\
US 2020 & U.S. Politics & 1.01 & [0.97, 1.04] & 0.97 & [0.92, 1.02] & 10.68 & 0.93 & 9.92 & 0.81 \\
Capitol Hill & U.S. Politics & 0.77 & [0.69, 0.86] & 0.86 & [0.8, 0.92] & 9.06 & 1.20 & 9.36 & 1.14 \\
COVID-19 & U.S. Politics & 0.96 & [0.92, 1.0] & 0.93 & [0.89, 0.97] & 10.04 & 0.94 & 10.00 & 1.03 \\
Black Lives Matter & U.S. Politics & 1.00 & [0.96, 1.04] & 0.94 & [0.9, 0.99] & 9.86 & 0.84 & 9.82 & 0.92 \\
NFL Draft & NFL & 0.87 & [0.81, 0.92] & 0.82 & [0.76, 0.87] & 9.89 & 0.62 & 9.11 & 0.66 \\
NFL Kickoff Game & NFL & 0.84 & [0.79, 0.88] & 0.76 & [0.71, 0.82] & 10.17 & 0.61 & 9.30 & 0.60 \\
SuperBowl LIV & NFL & 0.77 & [0.72, 0.82] & 0.82 & [0.76, 0.87] & 9.30 & 0.81 & 9.48 & 0.64 \\
NFL Trades & NFL & 0.82 & [0.76, 0.87] & 0.67 & [0.6, 0.73] & 9.73 & 0.58 & 8.98 & 0.67 \\
PlayOff & NFL & 0.79 & [0.74, 0.85] & 0.84 & [0.8, 0.89] & 9.60 & 0.67 & 10.29 & 0.63 \\
\hline
\end{tabular}
}
\end{table*}

\pagebreak[2]
\clearpage
\section*{Supplementary Information, Section 7: Wasserstein distance and Semantic Diversity}\label{sm_density_ws}
\begin{figure*}[h!]
    \centering
    \includegraphics[width=0.66\textwidth]{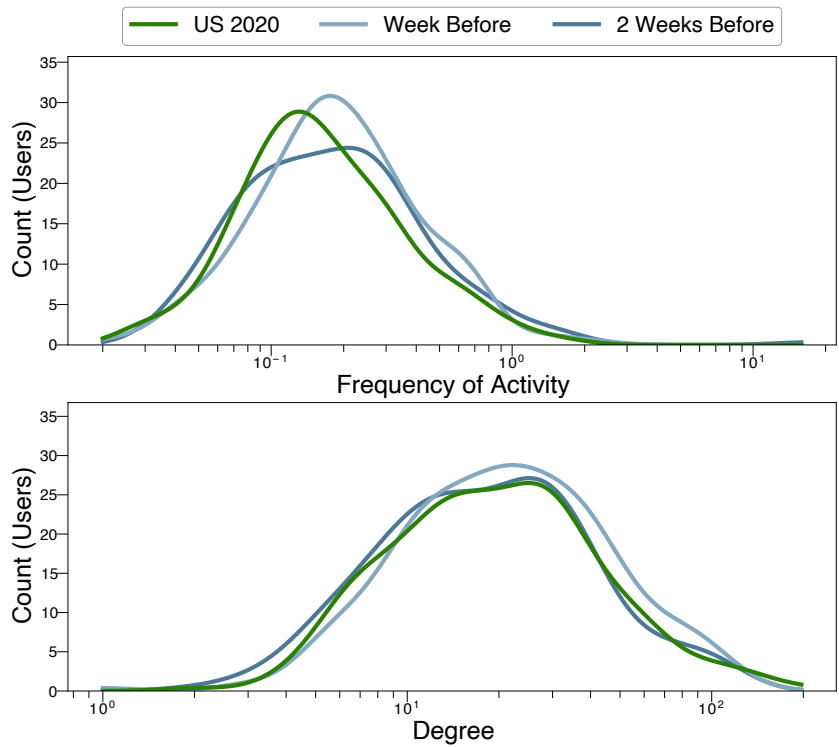}
    \caption{Histogram of the activity frequency (top panel) and the degree (bottom panel) of the European community, during the US 2020 election (green); week before (light blue) and two weeks before (blue).}
    \label{fig_sm_density_ws_eu}
\end{figure*}

We compare the weekly distributions of the activity frequency of the political communities during the US 2020 election (shared event) by computing the Wasserstein distance. In the European case, we find a low distance value during the exogenous event (0.03), comparable with the distance between the distributions of the week before and the two weeks before (0.08). As shown in Figure \ref{fig_sm_density_ws_eu}, the distributions are similar. On the contrary, in the American case we find a large value during the exogenous event (0.35); while the distributions of the week before and the two weeks before are similar (0.009) (See Figure). Similar values are obtained for the degree and the users' semantic diversity.
In Figure \ref{fig_sm_density_w2v_us} we report the peers and users' semantic diversity of the American community during the US 2020 election. 
The shift of peer distribution is stronger than that of users.
\begin{figure*}[h!]
    \centering
    \includegraphics[width=0.66\textwidth]{supplementary/density_ws/freq_degree_politics.pdf}
    \caption{Histogram of the activity frequency (top panel) and the degree (bottom panel) of the American community, during the US 2020 election (red); week before (light blue) and two weeks before (blue).}
    \label{fig_sm_density_ws_us}
\end{figure*}
\newpage
\begin{figure*}[h!]
    \centering
    \includegraphics[width=0.66\textwidth]{supplementary/density_ws/w2v_users_europe.pdf}
    \caption{Histogram of the peers' semantic diversity (top panel) and the users' semantic diversity (bottom panel) of the European community, during the US 2020 election (green); week before (light blue) and two weeks before (blue).}
    \label{fig_sm_density_w2v_eu}
\end{figure*}
\begin{figure*}[h!]
    \centering
    \includegraphics[width=0.66\textwidth]{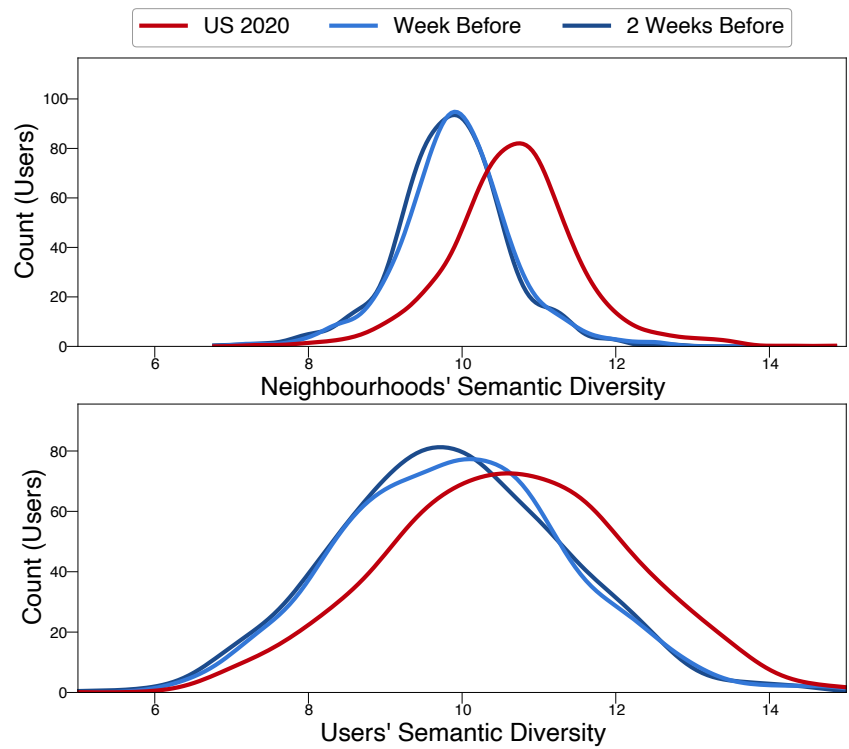}
    \caption{Histogram of the peers' semantic diversity (top panel) and the users' semantic diversity (bottom panel) of the American community, during the US 2020 election (green); week before (light blue) and two weeks before (blue).}
    \label{fig_sm_density_w2v_us}
\end{figure*}
\pagebreak[2]
\clearpage
\section*{Supplementary Information, Section 8: Post Displacements}\label{sm_displacements}
For each post we compute the average displacement in the semantic space, where we consider as displacement the euclidean distance between a comment and its following. We, then, consider only the posts where the dutiful users interact and compute for each post the user's average semantic diversity. Figure \ref{fig_sm_displacements} displays that the average semantic displacement of each post tends to increase as the semantic diversity of the user also increases. Moreover, we observe that the user's average semantic diversity is different for each community, as can be seen also from Figure \ref{fig_sm_joyplot}.
\begin{figure*}[h!]
    \centering
    \includegraphics[width=0.66\textwidth]{supplementary/displacement.pdf}
    \caption{It is shown the relation between the average semantic displacement and the average users' semantic diversity for the different communities (colors). The events are NBA Trades, NBA Restart, NBA Regular Season, US 2020 election, Capitol Hill, Trump Trial, NFL Kickoff Game, SuperBowl LIV, NFL Draft, Brexit. }
    \label{fig_sm_displacements}
\end{figure*}
\pagebreak[2]
\clearpage
\section*{Supplementary Information, Section 9: Compression Ratios and Heaps' Law}\label{sm_heap}
\begin{figure*}[h!]
    \centering
    \includegraphics[width=1.0\textwidth]{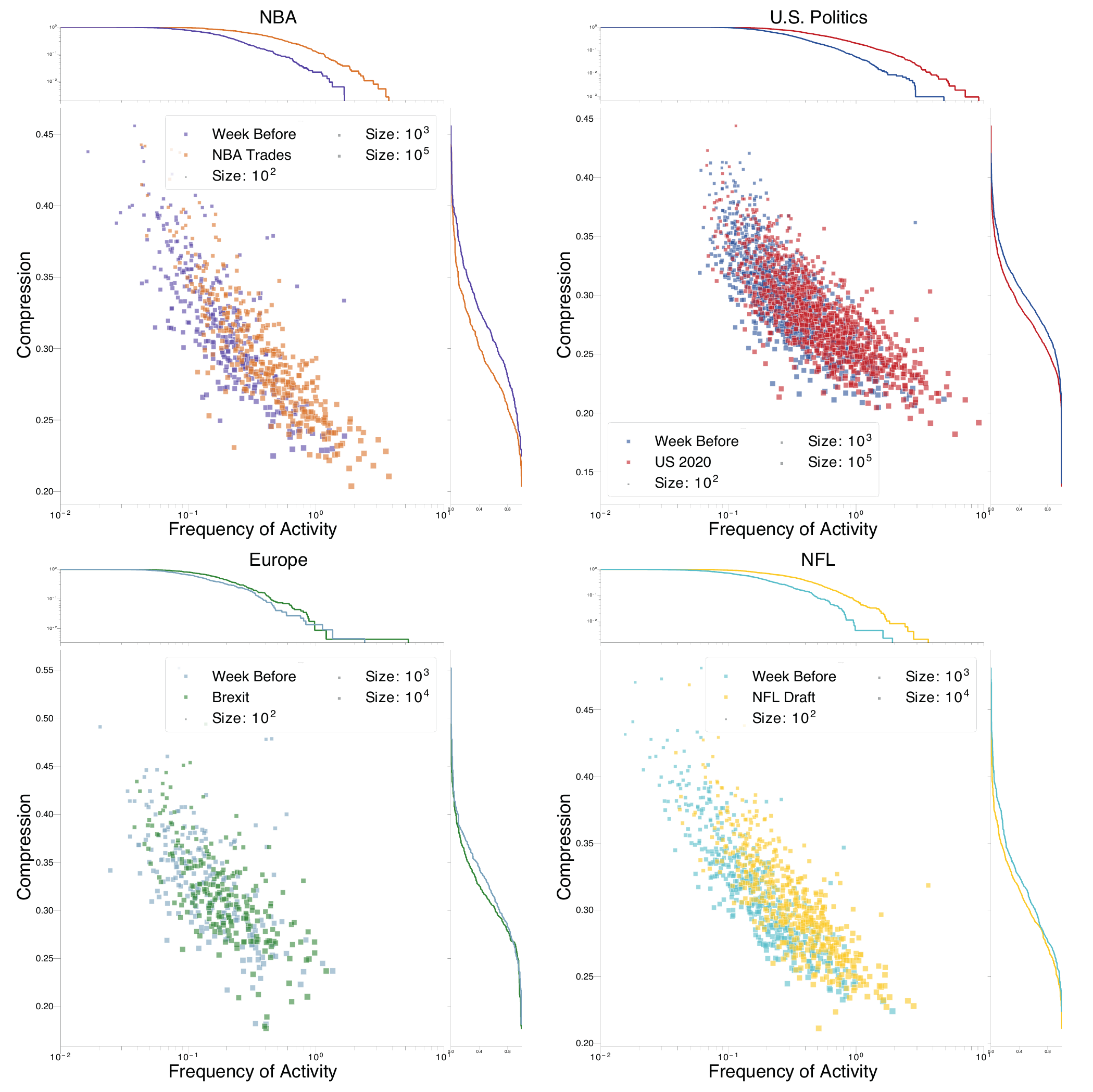}
    \caption{The central panels show the relationship between users' text compression and their frequency of activity for Europe (Brexit), NBA (NBA Trades), U.S. Politics (2020 election), and NFL (draft week). Each marker represents activity in a specific week for a user, with its size proportional to the total text size they produced that week. Marginal plots present survival functions of each variable.
    }
    \label{fig_sm_heap}
\end{figure*}

\pagebreak[2]
\clearpage
\section*{Supplementary Information, Section 10: Conversation changes in U.S. Politics and NBA Communities in June 2016}\label{sm_new_events}

\begin{figure*}[b!]
    \centering
    \includegraphics[width=\textwidth]{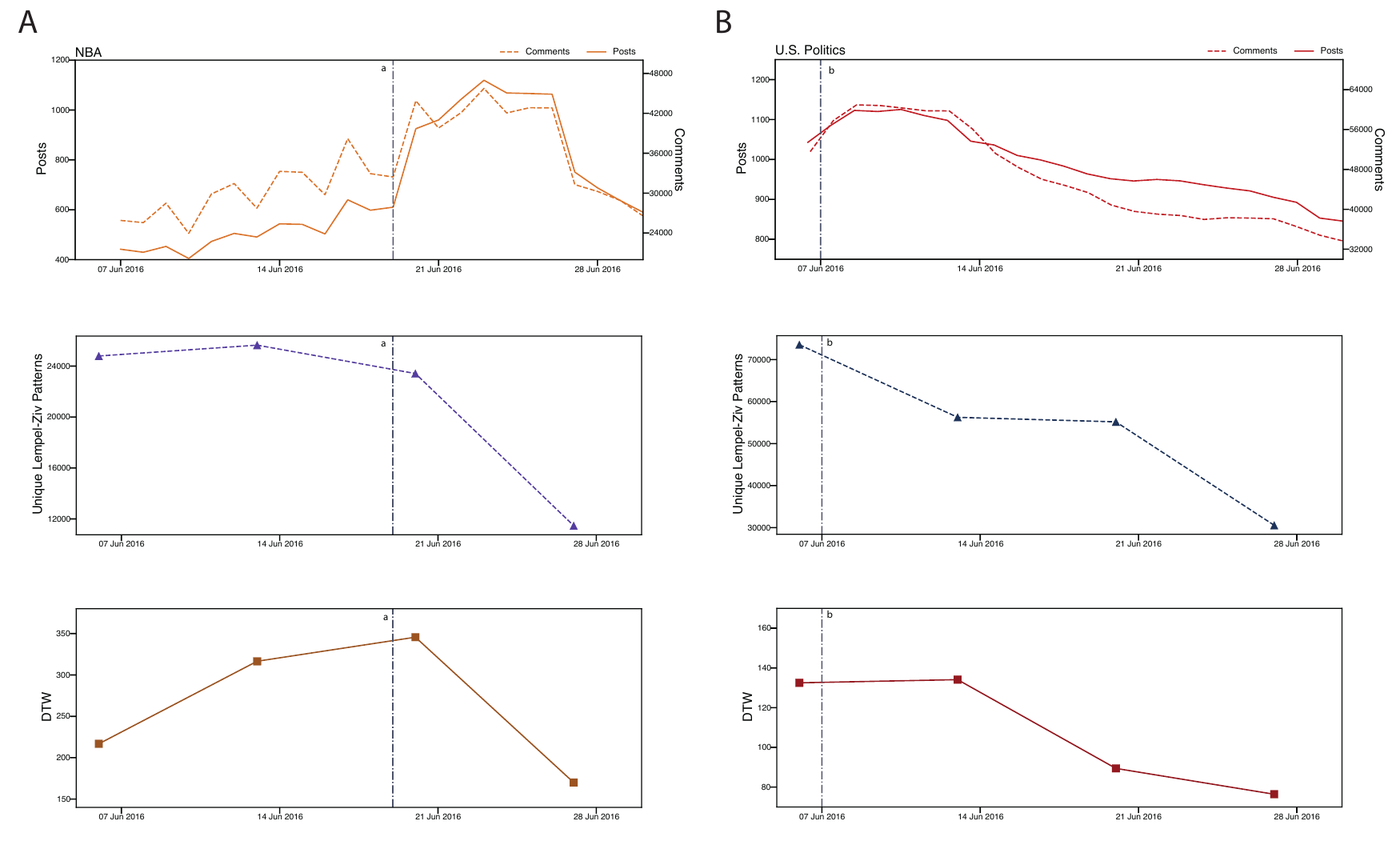}
    \caption{
    Community engagement and discourse Changes in U.S. Politics and NBA Communities in June 2016. 
    Panel A and Panel B represent the NBA and U.S. Politics communities respectively.
    The upper panels depict the post and comment activities around the events. 
    The middle panels illustrate the decrease in the number of unique patterns identified through the Lempel-Ziv scheme following the events. 
    The bottom panels show the peak in Dynamic Time Warping distance during the events, indicating significant shifts in discourse patterns. 
    }
    \label{fig_sm_new_events}
\end{figure*}

In this supplementary section, we present an extended analysis conducted on data from the U.S. politics (r/politics) and NBA (r/nba) communities during June 2016. 
This period was marked by two significant events: the emergence of the email controversy involving Hillary Clinton within the U.S. community on June 7, 2016, and the 2016 NBA Finals within the NBA community on June 19, 2016.
First, we notice a surge in community engagement around these events, as evidenced by the increased post and comment activities (refer to the upper panels of Figure \ref{fig_sm_new_events}).
Then, we applied two key metrics from the main text to capture shifts in community discourse: dynamic time warping distance and the identification of unique patterns via the Lempel-Ziv scheme. 
We observe that the number of unique patterns identified decreased following the events (refer to the middle panels of Figure \ref{fig_sm_new_events}).
Conversely, the dynamic time warping distance reached its peak during these events (refer to the bottom panels of Figure \ref{fig_sm_new_events}). 

\end{document}